\documentclass[trackchanges,twocolumn,resetfootnote]{aastex701}

\usepackage{comment}

\begin{document}
\title{The Galactic Phosphorus Survey. I. Chemical Evolution across the Galactic Disk from $\sim$750 FGK Stars
\footnote{Based on observations obtained with the Hobby-Eberly Telescope (HET), which is a joint project of the University of Texas at Austin, the Pennsylvania State University, Ludwig-Maximillians-Universitaet Muenchen, and Georg-August Universitaet Goettingen. The HET is named in honor of its principal benefactors, William P. Hobby and Robert E. Eberly.}}

\author[orcid=0000-0002-7587-7072]{Dionysios Gakis}
\affiliation{Department of Astronomy, The University of Texas at Austin, 2515 Speedway Boulevard, Austin, TX 78712, USA}
\email[show]{dgakis@utexas.edu}  

\author[orcid=0000-0002-1423-2174]{Keith Hawkins} 
\affiliation{Department of Astronomy, The University of Texas at Austin, 2515 Speedway Boulevard, Austin, TX 78712, USA}
\email{keithhawkins@utexas.edu}

\author[orcid=0000-0002-0475-3662]{Zachary G. Maas}
\affiliation{Department of Astronomy, Indiana University, Bloomington, IN 47405, USA}
\email{zmaas@indiana.edu}

\author[orcid=0000-0002-3456-5929]{Christopher Sneden} 
\affiliation{Department of Astronomy, The University of Texas at Austin, 2515 Speedway Boulevard, Austin, TX 78712, USA}
\email{chris@astro.as.utexas.edu}

\author[orcid=0000-0002-2516-1949]{Melike Afşar} 
\affiliation{Department of Astronomy, The University of Texas at Austin, 2515 Speedway Boulevard, Austin, TX 78712, USA}
\affiliation{Department of Astronomy and Space Sciences, Ege University, 35100 Bornova, İzmir, Turkey}
\email{melike.afsar@gmail.com}

\author[orcid=0000-0003-0595-5132]{Natalie R. Hinkel} 
\affiliation{Louisiana State University, Department of Physics and Astronomy, 202 Nicholson Hall, Baton Rouge, LA 70803, USA}
\email{natalie.hinkel@gmail.com}

\author[orcid=0000-0002-9768-2815]{Chenguang Sun} 
\affiliation{Department of Earth and Planetary Sciences, Jackson School of Geosciences, University of Texas at Austin, Austin, TX 78712, USA}
\email{csun@jsg.utexas.edu}

\author{Wenwei Liang} 
\affiliation{Department of Earth and Planetary Sciences, Jackson School of Geosciences, University of Texas at Austin, Austin, TX 78712, USA}
\email{wenweiliang666@utexas.edu}


\begin{abstract}

Phosphorus (P) is an odd-$Z$ element produced primarily in massive stars and is also important for planetary chemistry as an element essential to life. However, its evolution over cosmic time remains poorly constrained because precise and homogeneous stellar P measurements are scarce. We determine the relationship between [P/Fe] and [Fe/H] over a range of metallicities across the Galactic disk and assess its variation between different stellar populations using the largest uniformly analyzed sample to date, comprising $\sim$750 FGK stars with high-resolution near-infrared spectra. Stellar parameters are determined through a spectro-photometric approach and P abundances are derived by fitting synthetic spectra to the P~I 10529.52~\AA\ absorption feature. We confirm that [P/Fe] increases toward lower metallicities down to [Fe/H] $\sim$ -1.0 dex, where it appears to peak, broadly resembling the rise seen in $\alpha$-element abundance ratios over the same metallicity range and supporting a dominant contribution from massive stars on short core-collapse supernova timescales. When separating the sample into Galactic disk populations, thick-disk stars are systematically enhanced in [P/Fe] relative to the thin disk by $\sim$0.08 dex at fixed metallicity. We also measure a radial [P/Fe] gradient of $-0.124 \pm 0.006$ dex kpc$^{-1}$ and a vertical gradient of $+0.052 \pm 0.007$ dex kpc$^{-1}$. Although current chemical evolution models struggle to reproduce the observed trends, this survey increases the number of homogeneously measured stellar P abundances by approximately a factor of three, providing a new benchmark for Galactic chemical evolution models and the phosphorus inventory available to planetary systems.

\end{abstract}


\section{Introduction} 

Phosphorus (P) is an astrophysically and biochemically significant element \citep{hinkel2020influence} whose cosmic origin remains poorly constrained \citep{Hinkel2014,Maas2022}. As one of the bio-essential CHNOPS elements (carbon, hydrogen, nitrogen, oxygen, phosphorus, and sulfur), P is fundamental to life as we know it, forming a key component of DNA, RNA, and other biologically important molecules \citep{Schlesinger2013}. Despite its importance, the nucleosynthetic pathways that can produce P are still uncertain, and its behavior across different Galactic populations remains relatively unconstrained compared to other elements. More homogeneous P abundance measurements across a broad range of metallicities and Galactic components are therefore needed to constrain its stellar production and clarify how P enrichment proceeds alongside that of other major elements.

The first stellar P abundance study was carried out by \citet{Caffau2011}, who analyzed high-resolution Y-band spectra of four FGK dwarfs and found a declining [P/Fe] trend with increasing [Fe/H], suggesting an $\alpha$-element-like behavior. Using near-UV P I lines observed with Hubble Space Telescope - Space Telescope Imaging Spectrograph, \citet{Jacobson2014} and \citet{Roederer2014} extended P measurements to lower metallicities, deriving abundances for 13 metal-poor stars down to [Fe/H] $\sim -3.8$ and finding [P/Fe] ratios near solar for [Fe/H] $\lesssim -1.5$. Additional HST observations by \citet{Spite2017} provided a P measurement for an extremely metal-poor star at [Fe/H] $= -2.25$, with [P/Fe] $= -0.32$.

Near-infrared spectroscopy, primarily using Y-band P I lines, soon became the dominant approach for stellar P measurements. Studies of disk dwarfs and giants confirmed a declining [P/Fe] trend with increasing [Fe/H] \citep{Caffau2016,Maas2017,Afar2018,Maas2019,BocekTopcu2019,Caffau2019,Bocek2020,Sneden2021,Sadakane2022,Ozdemir2025,Matsunaga2026}. \citet{Maas2022} measured P in 163 FGK stars using the 10529.52~\AA\ line and found enhanced [P/Fe] in the thick disk relative to the thin disk at similar metallicity\footnote{Throughout this work, we use “metallicity” as shorthand for the stellar iron abundance, [Fe/H], unless otherwise stated, while recognizing that [Fe/H] does not necessarily trace the abundances of other heavy elements.}, consistent with an $\alpha$-element-like enrichment pattern. In the H-band, \citet{Nandakumar2022} measured P in nearby K giants and found that P behaves more similarly to $\alpha$-elements than to neighboring odd-$Z$ elements, in agreement with previous literature trends. More recently, \citet{Jian2025} and \citet{Catanzaro2025} measured P in open cluster stars and Cepheids, finding trends consistent with Galactic chemical evolution expectations; \citet{Jian2025} also reported a [P/Fe]--age relation among open clusters older than 1~Gyr.



Large spectroscopic surveys have further advanced the study of stellar P by enabling measurements across thousands of stars and diverse Galactic environments. Using APOGEE H-band spectra, \citet{Hawkins2016} reanalyzed approximately 2000 red giants and found a decreasing [P/Fe] trend with increasing metallicity. However, P abundances in APOGEE rely on blended features in a molecular-rich spectral region and are affected by telluric contamination, leading to larger systematic uncertainties compared to targeted high-resolution studies \citep{Johnson2020}. Despite these limitations, APOGEE has proven particularly powerful for identifying rare populations. Within APOGEE, \citet{Masseron2020} identified 15 P-rich ([P/Fe] $>+1.2$) stars showing enhancements in Mg, Si, Al, and $s$-process elements, suggesting a distinct nucleosynthetic origin. This population was later expanded by \citet{Brauner2023}, who analyzed 87 APOGEE DR17 stars and found that 78 exhibit enhancements ([P/Fe] $>+0.8$), establishing P-rich stars as a distinct class.

P abundances in the Galactic bulge have also been investigated using APOGEE spectra. \citet{Barbuy2025} studied 58 metal-poor bulge stars and found that approximately one-third are moderately P-rich ([P/Fe] $>+0.45$), with a peak near [Fe/H] $\sim -1.0$. Extending this work to higher metallicities, \citet{Ernandes2026} analyzed 78 metal-rich bulge spheroid stars ([Fe/H] $>-0.8$) and reported a clear P enhancement around [Fe/H] $\sim -0.7$, reinforcing the presence of a distinctive chemical signature in the early bulge population. Finally, \citet{Barbuy2025b} identified moderately P-enhanced stars in two bulge globular clusters using APOGEE spectra, while \citet{Camargo2026} extended this work to the bulge clusters NGC 6539 and NGC 6569, finding one clearly P-rich star and evidence for more moderate P enhancement, respectively. 


The observed Galactic [P/Fe] trends provide important constraints on the nucleosynthetic origin of P. Current models indicate that P is mainly synthesized in massive stars, where $^{31}$P is produced through neutron capture on $^{30}$Si followed by $\beta^-$ decay during hydrostatic carbon and neon shell burning, and later expelled during core-collapse supernovae \citep{Woosley1995,Clayton2003}. Enhanced P abundances have been observed in the supernova remnant Cas A \citep{Koo2013}, and the similarity of Galactic [P/Fe] trends to those of $\alpha$-elements such as Mg and Si further supports a massive-star origin \citep{Maas2022}. However, theoretical yields must be boosted by factors of two to three to reproduce observations \citep{Cescutti2012}. Type Ia SNe and AGB stars are thought to contribute negligibly \citep{Travaglio2004,Karakas2016}, though more exotic channels—such as mergers between the oxygen- and carbon-burning shells in rotating massive stars, which can modify the late-stage nucleosynthesis conditions—have been proposed as possible sources of additional P \citep{Ritter2018}. 

This discrepancy between the predicted massive-star yields and the observed P abundances becomes even more pronounced in the class of P-rich stars \citep{Masseron2020,Brauner2023,Brauner2024}, which show strong P enhancements alongside O, Mg, Si, Al, Ce and s-process elements. 
Their abundance patterns point to rare nucleosynthetic events rather than binary mass transfer, with possibilities including convective-reactive burning or intermediate neutron-capture (i-process) in massive stars. Novae, particularly ONe novae, have also been suggested as contributors at intermediate metallicities \citep{Bekki2024}, though they cannot yet reproduce the full chemical signature of P-rich stars. Expanding the census of P measurements is thus critical to establish accurate [P/Fe]–[Fe/H] relations and to evaluate the role of metallicity, stellar type, and kinematic population in shaping galactic enrichment. 

In this work, we 
perform a uniform study of P abundances in the Galactic disk with near-infared spectra of FGK stars. 
By deriving homogeneous P abundances for a well-characterized sample, we investigate how P behaves across the thin and thick disks, and compare the observed trends with theoretical yields to constrain the dominant nucleosynthetic sources of P in the Milky Way.

The remainder of this paper is organized as follows. Section \ref{sec:sec2} describes the observations and data reduction. Section \ref{sec:sec3} details the derivation of stellar parameters and abundance analysis. Section \ref{sec:sec4} presents the results and model comparisons and Section \ref{sec:sec5} summarizes the conclusions and outlines prospects for future.

\section{Data properties}  \label{sec:sec2}

\subsection{Stellar sample} \label{sec:sec2.1}

The spectroscopic data analyzed in this work were obtained with the Habitable-zone Planet Finder (HPF), a fiber-fed, near-infrared echelle spectrograph mounted on the 10\,m Hobby--Eberly Telescope \citep{HET1,HET2}. HPF operates over a wavelength range of $0.8$--$1.3~\mu$m at a resolving power of $R \approx 50{,}000$ \citep{Mahadevan2012,Mahadevan2014}. This spectral coverage includes the P I absorption feature at 10529.52~\AA, which serves as the primary diagnostic for P abundance measurements in this work. HPF’s high spectral resolution and Y-band coverage make it uniquely well suited for detecting the intrinsically weak P I feature.

Our stellar sample is drawn from previous high-resolution HPF studies by \citet{Sneden2022} and \citet{Mallick2025}, whose spectra we reanalyze using a uniform methodology. The final sample from these two studies comprises 601 stars. These earlier works did not include P abundance determinations. We further augment the sample with 154 stars also from HPF from \citet{Maas2022}, who measured P abundances for these stars; however, we reprocess their spectra within our framework to ensure that the stellar parameters and P abundances for the full sample are derived homogeneously. The combined dataset comprises a total of 755 FGK stars spanning both dwarf and giant evolutionary stages. The signal-to-noise ratio (S/N) of the final spectra ranges from $\sim$ 80-350/pixel in the vicinity of the P I feature.

All raw exposures were reduced using the \texttt{Goldilocks} pipeline\footnote{\url{https://github.com/grzeimann/Goldilocks_Documentation}}, which is optimized for HPF data. The reduction procedure includes reference pixel-based bias correction, non-linearity correction, scattered light subtraction, flat-fielding, optimal extraction of spectral orders, and wavelength calibration \citep{Ninan2018}. The pipeline produces wavelength-calibrated spectra for science, sky, and calibration fibers, along with corresponding uncertainty arrays that are propagated through each reduction step. No telluric correction is applied at this stage to avoid introducing additional noise or systematic effects that could impact weak spectral features such as the P I line. The reliability of the \texttt{Goldilocks} reduction for precision abundance studies on P has been demonstrated in previous works \citep[e.g.,][]{Sneden2021,Maas2022}.

The reduced spectra were further processed using the \texttt{muler} package\footnote{\url{https://github.com/OttoStruve/muler}}. This processing includes masking of bad pixels, subtraction of the sky spectrum, blaze function correction, merging of individual echelle orders, and continuum normalization. Radial velocities were determined via cross-correlation with a synthetic high-resolution template computed at representative median atmospheric parameters of $T_{\rm eff}=4900$~K, $\log g=2.2$, and $[\mathrm{Fe/H}]=-1.0$, using \texttt{iSpec} \citep{Blanco-Cuaresma2014,Blanco-Cuaresma2019}, and all spectra were shifted to the stellar rest frame prior to combination. For each target, multiple exposures were co-added using inverse-variance weighting to maximize the signal-to-noise ratio.

\subsection{Supplemental stellar data} \label{sec:sec2.2}

We supplement the HPF spectra with external astrometric, photometric, and kinematic information to determine stellar parameters and Galactic population membership. We adopt Gaia Data Release 3 (DR3; \citealp{GaiaDr3}) astrometry and apply the parallax zero-point correction following the prescription of \citet{Lindegren2021}. Distances are preferentially taken from the photogeometric Gaia DR3 distance catalog of \citet{Bailer-Jones2021}; when photogeometric distances are unavailable, we adopt the corresponding geometric distances. Photometric inputs include near-infrared $J$, $H$, and $K_s$ magnitudes from 2MASS \citep{2MASS:2003,2MASS:2006} and optical $G$, $G_{\mathrm{BP}}$, and $G_{\mathrm{RP}}$ magnitudes from Gaia \citep{Gaia,GaiaDr3}. For Gaia photometry, we apply the bright-end correction described by \citet{Riello2021} to $G$, $G_{\mathrm{BP}}$, and $G_{\mathrm{RP}}$ before correcting the magnitudes for extinction.


Interstellar reddening, $E(B-V)$, is obtained from the Bayestar 3D dust map \citep{Green2019} at the Gaia-corrected distance. Band-dependent extinction coefficients, $R_\lambda$, are computed using the \texttt{extinction\_coefficient} package\footnote{\url{https://github.com/vnohhf/extinction_coefficient?tab=readme-ov-file}} \citep{Zhang2023}, accounting for the stellar effective temperature (or in practice, $G_{BP}-G_{RP}$ color) and filter response. The extinction in each photometric band, $A_\lambda$, is then calculated as
\begin{equation}
A_\lambda = R_\lambda E(B-V),
\end{equation}
and all magnitudes are corrected for extinction prior to their use in the stellar-parameter inference (Section \ref{sec:sec3.1}).

Galactocentric positions and velocities are derived from Gaia DR3 astrometry and radial velocities together with the adopted distance estimates. The transformation is performed with \texttt{astropy} \citep{2013A&A...558A..33A} in a right-handed Galactocentric Cartesian frame, adopting a Sun--Galactic center distance of $R_\odot=8.122$~kpc \citep{Gravity2018}, a solar height of $z_\odot=20.8$~pc \citep{BennettBovy2019}, and a total solar velocity relative to the Galactic center of $(v_{x,\odot},v_{y,\odot},v_{z,\odot})=(12.9,245.6,7.78)$~km~s$^{-1}$. We denote the resulting Galactocentric Cartesian coordinates by $(x,y,z)$. In this frame, the positive $x$, $y$, and $z$ directions point approximately toward the Galactic center, in the direction of Galactic rotation, and toward the North Galactic Pole, respectively. We define $R_{\rm Gal}=(x^2+y^2)^{1/2}$ as the projected distance from the Galactic center in the disk plane and $z_{\rm Gal}=z$ as the vertical displacement from the Galactic midplane.

Space velocities relative to the local standard of rest are calculated separately and denoted by $(U_{\rm LSR}, V_{\rm LSR}, W_{\rm LSR})$. We adopt the convention that $U_{\rm LSR}$ is positive toward the Galactic center, $V_{\rm LSR}$ is positive in the direction of Galactic rotation, and $W_{\rm LSR}$ is positive toward the North Galactic Pole. The transformation uses the solar peculiar motion $(U_\odot,V_\odot,W_\odot)=(11.1, 12.24, 7.25)$~km~s$^{-1}$ from \citet{Schonrich2010}.

To propagate the kinematic uncertainties, we generate 2000 Monte Carlo realizations for each star. In each realization, the adopted distance, proper motions, and Gaia radial velocity are independently drawn from Gaussian distributions centered on their measured values, with standard deviations equal to their corresponding uncertainties. Each realization is transformed to both the Galactocentric and LSR frames, and the means and standard deviations of the resulting position and velocity distributions are adopted as the final values, $U_{\rm LSR}$, $V_{\rm LSR}$, and $W_{\rm LSR}$, and uncertainties, $\sigma_U$, $\sigma_V$, and $\sigma_W$, respectively. 

Velocities are used to compute probabilistic assignments to the thin disk, thick disk, and halo populations ($P_{\rm thin}$, $P_{\rm thick}$, and $P_{\rm halo}$, respectively), following the formalism of \citet{Ramirez2013} and \citet{Bensby2014}, which assumes Gaussian velocity distributions for each Galactic component. The corresponding likelihood ratios are evaluated using the analytic expressions summarized in Appendix~A of \citet{Hackshaw2024}. The derived Galactocentric positions, space velocities and uncertainties, and membership probabilities for all stars are listed in Table~\ref{table::gal_kinematics_positions}.


\begin{figure}
    \centering
\includegraphics[width=0.45\textwidth]{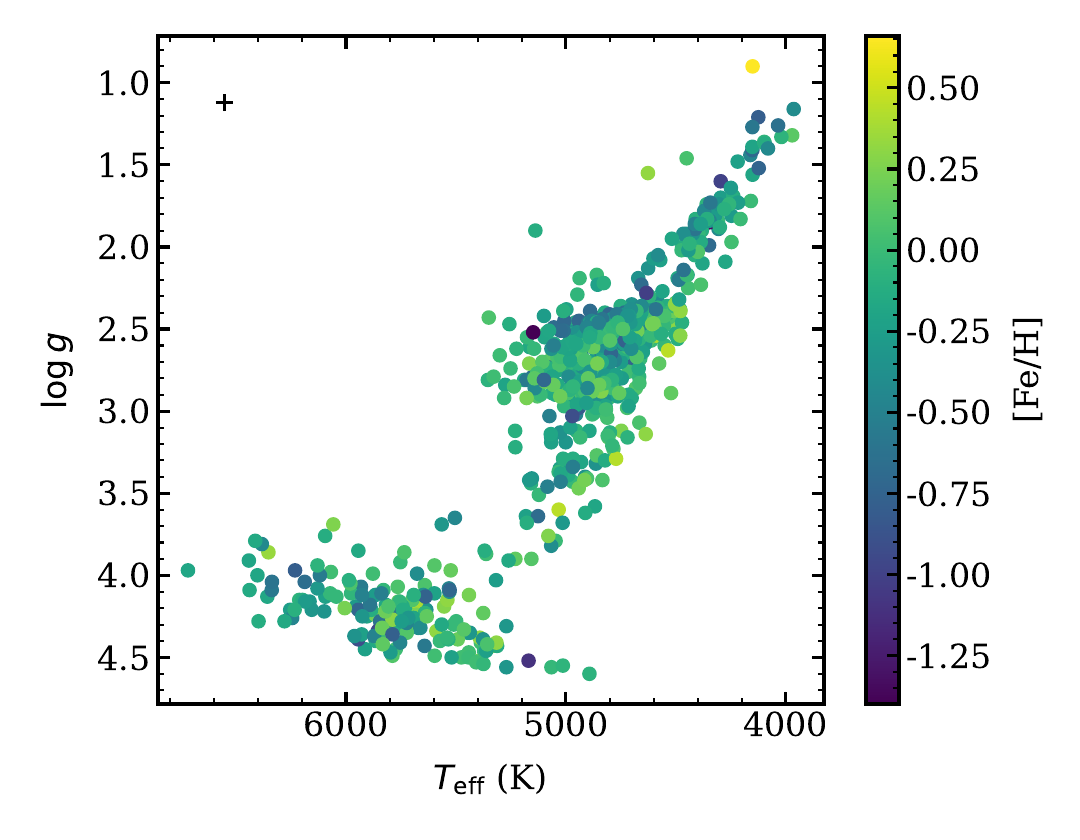}
\caption{Kiel diagram of the stellar sample analyzed in this work, showing surface gravity ($\log g$) as a function of effective temperature ($T_{\rm eff}$), with points color-coded by spectroscopic metallicity ($[\mathrm{Fe/H}]$). The black error bars in the upper-left corner indicate the mean uncertainties in $T_{\rm eff}$ and $\log g$.}   
\label{fig:kiel}
\end{figure}

\begin{deluxetable*}{l c c c c c c c c c c c}
\tablewidth{0pt}
\tabletypesize{\footnotesize}
\tablecaption{Galactocentric Positions, Space Velocities, and Population Membership Probabilities}
\label{table::gal_kinematics_positions}
\tablehead{
\colhead{Name} &
\colhead{$R_{\rm Gal}$} & \colhead{$z_{\rm Gal}$} &
\colhead{$U_{\rm LSR}$} & \colhead{$\sigma_U$} &
\colhead{$V_{\rm LSR}$} & \colhead{$\sigma_V$} &
\colhead{$W_{\rm LSR}$} & \colhead{$\sigma_W$} &
\colhead{$P_{\rm thin}$} & \colhead{$P_{\rm thick}$} & \colhead{$P_{\rm halo}$} \\
\colhead{} &
\colhead{(kpc)} & \colhead{(kpc)} &
\colhead{(km s$^{-1}$)} & \colhead{(km s$^{-1}$)} &
\colhead{(km s$^{-1}$)} & \colhead{(km s$^{-1}$)} &
\colhead{(km s$^{-1}$)} & \colhead{(km s$^{-1}$)} &
\colhead{} & \colhead{} & \colhead{}
}
\startdata
2MASS J00230934+0007429 & 8.24 & -0.67 & 47.5 & 0.5 & -36.7 & 0.5 & 20.6 & 0.4 & 0.95 & 0.05 & 0.00 \\
2MASS J03065740+5449428 & 10.20 & -0.11 & 42.7 & 1.8 & -5.6 & 1.4 & -15.8 & 1.4 & 0.99 & 0.01 & 0.00 \\
2MASS J03221188+2736271 & 8.30 & -0.07 & 1.4 & 0.1 & 1.6 & 0.3 & -3.9 & 0.2 & 0.99 & 0.01 & 0.00 \\
2MASS J04383875-0258161 & 10.13 & -1.22 & -49.6 & 0.9 & 8.5 & 0.5 & 4.7 & 0.9 & 0.99 & 0.01 & 0.00 \\
2MASS J05524768+4422415 & 9.01 & 0.17 & 29.8 & 0.1 & -6.2 & 0.3 & 5.3 & 0.1 & 0.99 & 0.01 & 0.00 \\
$\cdots$ & $\cdots$ & $\cdots$ & $\cdots$ & $\cdots$ & $\cdots$ &
$\cdots$ & $\cdots$ & $\cdots$ & $\cdots$ & $\cdots$ & $\cdots$ \\
\enddata
\tablecomments{(This table is available in its entirety in machine-readable form.)}
\end{deluxetable*}

\section{Atmospheric parameters \& abundances} \label{sec:sec3}

\subsection{Atmospheric parameters} \label{sec:sec3.1}

Deriving accurate atmospheric parameters from HPF spectra is challenging because traditional excitation and ionization balance techniques based on Fe I and Fe II lines are poorly suited to the Y-band (8,000–13,000~\AA), where only a handful of Fe II lines are detectable, particularly in cool FGK giants \citep{kondo2019,Sneden2021}. To overcome these limitations, we adopted a hybrid spectro-photometric approach, which combines photometry, parallaxes, and spectral fitting of Fe~I lines in an iterative scheme, broadly similar to other Bayesian spectro-photometric frameworks such as \texttt{MINESweeper} \citep{Cargile2020minesweeper}.

In short, our procedure consists of a photometric step followed by a spectroscopic one. In the photometric step, effective temperatures ($T_{\rm eff}$) are first determined using empirical color-$T_{\rm eff}$ calibrations. Surface gravities ($\log g$) are constrained using Gaia astrometry and $K_s$-band photometry through fundamental relations. Microturbulent velocities ($\xi$) are prescribed from empirical relations. In the spectroscopic step, metallicities ([Fe/H]) and rotational velocities ($v\sin i$) are determined iteratively through spectral fitting of Fe~I lines fixing the $T_{\rm eff}$, $\log g$, with $T_{\rm eff}$, $\log g$, and $\xi$ fixed to the values obtained in the photometric step.

More specifically, effective temperatures are estimated from empirical color--temperature relations calibrated for FGK stars by \citet{Mucciarelli2021}, using all available dereddened colors spanning optical to near-infrared combinations when possible. 
As \citet{Mucciarelli2021} provide separate calibrations for dwarfs and giants, we compute both solutions and adopt the branch whose final surface gravity is self-consistent, classifying stars with $\log g<3.5$ as giants and those with $\log g\geq3.5$ as dwarfs.
For each applicable color index, $C_i$, we obtain an independent temperature estimate, $T_{\mathrm{eff},i}$, explicitly accounting for the metallicity dependence in the \citet{Mucciarelli2021} relations. 
The final $T_{\mathrm{eff}}$ is obtained by combining these individual color-based estimates using inverse-variance weighting, where the variance for each relation includes both the calibration scatter and the photometry- and $[\mathrm{Fe}/\mathrm{H}]$-propagated uncertainties. Uncertainties in $T_{\mathrm{eff}}$ are propagated via Monte Carlo sampling of the input photometry and metallicity, and therefore capture the combined effects of measurement errors and the intrinsic scatter of the calibrations \citep{Mucciarelli2021}. This procedure is repeated as updated metallicities become available from the spectroscopic analysis.

Surface gravities are determined following the process outlined in Section 5 of GALAH DR4\footnote{\url{https://github.com/svenbuder/GALAH_DR4}} \citep{GalahDR4} which ties $\log g$ to Gaia-based luminosities, rather than fitting $\log g$ purely from spectral lines. Because near-infrared magnitudes are less sensitive to (often uncertain) extinction corrections than optical bands, we adopt the $K_s$ band to derive bolometric magnitudes and luminosities. Using the extinction-corrected magnitude $K_{s,0}$ and the adopted Gaia DR3 distance $D$, the bolometric magnitude is computed as in \cite{GalahDR4}:
\begin{equation}
M_{\rm bol} = K_{s,0} + 5 - 5\log(D) + BC(K_{s,0}),
\end{equation}
where $BC(K_{s,0})$ is the bolometric correction. 
We interpolate bolometric corrections from the \citet{Casagrande2018} tables as a function of $T_{\rm eff}$, $\log g$, and metallicity. In the first photometric step, the calculation is initialized separately for the dwarf and giant branches using $\log g=4.4$ and $2.5$, respectively. These initial values are used to evaluate the first bolometric correction and luminosity. Because $BC(K_s)$ depends explicitly on $\log g$, we then iteratively recompute $BC(K_{s,0})$, luminosity, stellar mass, and $\log g$ until convergence, typically within a few iterations.

Stellar masses ($\mathcal{M}$) and ages ($\tau$) are inferred simultaneously with surface gravities using PARSEC+COLIBRI isochrones \citep{Bressan2012,Marigo2017}, following the likelihood-based framework adopted in GALAH DR4 \citep[e.g.,][]{Lin2018}. We use the default PARSEC+COLIBRI grids adopted by GALAH DR4, with logarithmic ages spanning $\log(\tau/{\rm yr}) = 6.19$ to $10.17$ in steps of $0.01$. The metallicity grid consists of a coarse metal-poor component, $[\mathrm{M}/\mathrm{H}] = -2.75$ to $-0.75$ dex in steps of $0.25$ dex, and a finer metal-rich component, $[\mathrm{M}/\mathrm{H}] = -0.60$ to $+0.70$ dex in steps of $0.10$ dex. For each isochrone point, we compute a Gaussian likelihood using the differences between the current estimates and model values of $T_{\mathrm{eff}}$, $\log g$, $\log L$, and $[\mathrm{M}/\mathrm{H}]$, scaled by their uncertainties. Masses and ages are computed as likelihood-weighted averages over the retained isochrone points, and the inferred mass is used together with the luminosity and effective temperature to update the surface gravity iteratively. For the isochrone and bolometric-correction steps, we convert [Fe/H] to the global metallicity [M/H] because the PARSEC+COLIBRI isochrone tables and bolometric-correction grids are tabulated in [M/H]. As in GALAH DR4, we use the prescription of \citet{Salaris2005}, $[\mathrm{M}/\mathrm{H}] = [\mathrm{Fe}/\mathrm{H}] + \log \left(0.694 \cdot 10^{[\alpha/\mathrm{Fe}]} + 0.306\right)$, adopting an $\alpha$-enhancement that varies smoothly with [Fe/H] ([$\alpha$/Fe]$=0.4$ at [Fe/H]$<-1$, [$\alpha$/Fe]$=0.0$ at [Fe/H]$>0$, and linear interpolation in between). We adopt an initial value of $[\mathrm{Fe}/\mathrm{H}]=0.0\pm0.2$ and replace it in subsequent photometric--spectroscopic iterations with the [Fe/H] obtained from the preceding Fe~I line fit. This mapping places the stars on the [M/H] axis of the PARSEC+COLIBRI isochrone tables and bolometric-correction grid while allowing the spectroscopic analysis to remain expressed in [Fe/H].


The final surface gravity is then computed from
\begin{eqnarray}
    \log g = \log g_\odot + \log\left(\frac{\mathcal{M}}{\mathcal{M}_\odot}\right) \nonumber\\ + \ 4\log\left(\frac{T_{\rm eff}}{T_{{\rm eff},\odot}}\right)  - \log\left(\frac{L}{L_\odot}\right),
\end{eqnarray}
with $\log g_\odot = 4.438$, $T_{\mathrm{eff},\odot} = 5772~{\rm K}$, $\mathcal{M}$ expressed in solar units, and $\log(L/L_\odot)=-0.4(M_{\mathrm{bol}}-M_{\mathrm{bol},\odot})$ with $M_{\mathrm{bol},\odot}=4.75$. This relation combines the isochrone-based mass estimate with the photometrically inferred luminosity and effective temperature. 


The microturbulent velocity $\xi$ is not fit spectroscopically but is instead prescribed as a function of the atmospheric parameters inferred. For dwarf stars ($\log g > 3.5$), microturbulence is computed using the empirical calibration implemented in the \texttt{BACCHUS} spectral-analysis code \citep{bacchus} which depends on both effective temperature and surface gravity. For $T_{\rm eff} > $ 5250~{\rm K}:
\begin{eqnarray}
\xi_{\rm dwarf} = 1.15 + 2.0\times10^{-4}\,(T_{\rm eff}-5500) \nonumber\\
+ 3.95\times10^{-7}\,(T_{\rm eff}-5500)^2 
- 0.13\,(\log g-4.0) \nonumber\\
+ 0.13\,(\log g-4.0)^2 ,
\end{eqnarray}
whereas for $T_{\rm eff} < $ 5250~{\rm K}:
\begin{eqnarray}
\xi_{\rm dwarf} = 1.15 + 2.0\times10^{-4}\,(5250-5500) \nonumber\\
+ 3.95\times10^{-7}\,(5250-5500)^2 
- 0.13\,(\log g-4.0) \nonumber\\
+ 0.13\,(\log g-4.0)^2 .
\end{eqnarray}
This prevents the dwarf polynomial from being extrapolated to cooler temperatures, where the quadratic term would otherwise cause $\xi$ to increase again toward lower $T_{\mathrm{eff}}$.
For giant stars ($\log g \le 3.5$), we adopt the empirical formula of APOGEE DR13 \citep{APOGEEDR13}, also used by \citet{Maas2022}, expressed as a cubic polynomial of $\log g$,
\begin{eqnarray}
\log\xi_{\rm giant}
= 0.225
- 0.0228\,\log g \nonumber\\
+ 0.0297\,(\log g)^2 
- 0.0113\,(\log g)^3 .
\end{eqnarray}

\begin{deluxetable*}{cccc|cccc}
\tablewidth{0pt}
\tabletypesize{\footnotesize}
\tablecaption{Line selection \label{table::linelist}}
\tablehead{
\colhead{Wavelength (Å)} & \colhead{Species} & \colhead{$\chi$ (eV)} & \colhead{$\log(gf)$} & 
\colhead{Wavelength (Å)} & \colhead{Species} & \colhead{$\chi$ (eV)} & \colhead{$\log(gf)$}  
}
\startdata
8327.057  & Fe I & 2.196 & -1.550  & 10167.469 & Fe I & 2.196 & -3.930  \\
8387.773  & Fe I & 2.174 & -1.510  & 10216.314 & Fe I & 4.733 & -0.130  \\
8514.069  & Fe I & 2.196 & -2.200  & 10218.408 & Fe I & 3.069 & -2.760  \\
8674.746  & Fe I & 2.829 & -1.850  & 10340.886 & Fe I & 2.196 & -3.360  \\
8688.624  & Fe I & 2.174 & -1.200  & 10395.796 & Fe I & 2.174 & -3.210  \\
8824.221  & Fe I & 2.196 & -1.540  & 10881.758 & Fe I & 2.843 & -3.690  \\
9889.035  & Fe I & 5.029 & -0.450  & 11119.796 & Fe I & 2.843 & -2.200  \\
10145.561 & Fe I & 4.792 & -0.280  & 10529.524 & P I  & 6.949 &  0.240  \\
\enddata
\tablecomments{Wavelengths are given in air in units of Å, $\chi$ is the lower-level excitation potential in eV, and $\log(gf)$ is the logarithm of the product of the statistical weight $g$ and oscillator strength $f$. \label{table::linelist}}

\end{deluxetable*}

A set of 15 Fe I lines within the HPF Y-band coverage was selected for deriving metallicities (Table~\ref{table::linelist}), some of which have also been employed in previous HPF abundance analyses \citep{Sneden2021,Maas2022}. The starting Y-band line list was generated with the \texttt{linemake} code\footnote{\url{https://github.com/vmplacco/linemake}} \citep{linemake}. We then retained only lines that satisfied three practical criteria: they produced a measurable Fe I absorption feature in high-S/N observed spectra, their line cores and wings were well reproduced by synthetic spectra at representative stellar parameters, and they were not strongly affected by neighboring atomic or molecular blends within the fitted window. Candidate lines were checked by synthesizing spectra over a range of representative metallicities and verifying that the Fe I feature remained identifiable, minimally blended, and well reproduced across the relevant parameter space. Lines with poorly matched profiles, severe blending, uncertain continuum placement, or inconsistent behavior across stars were excluded from the final Fe line set.



For the spectroscopic step, all selected Fe lines (Table~\ref{table::linelist}) are analyzed simultaneously with $[\mathrm{Fe}/\mathrm{H}]$ and $v\sin i$ as free parameters using Bayesian inference, following similar Bayesian synthetic spectrum fitting approaches in the literature \citep[e.g.,][]{Gill2018,Tabernero2022}. Directly generating a new synthetic spectrum at every likelihood evaluation would be computationally expensive. We therefore constructed and trained a neural network spectral emulator to rapidly reproduce the synthetic spectra as a continuous function of the atmospheric parameters. 

The emulator is trained on a uniform grid of high-resolution synthetic spectra ($R\sim53{,}000$) generated in the $Y$ band with \texttt{Turbospectrum}, spanning $T_{\mathrm{eff}}=3600$--7000~K in steps of 200~K, $\log g=0.5$--5.0 in steps of 0.5~dex, $[\mathrm{Fe}/\mathrm{H}]=-2.5$--1.0 in steps of 0.5~dex, and $\xi=0.0$--3.0~km~s$^{-1}$ in steps of 1.0~km~s$^{-1}$. The synthetic spectra were randomly divided into a training set containing 70\% of the grid and a validation set containing the remaining 30\%. Only the training spectra were used to adjust the parameters of the neural network; the validation spectra were held out entirely during this optimization and used only to evaluate how accurately the trained emulator reproduced spectra that it had not been shown during training.

The emulator is a fully connected feed-forward neural network, meaning that information passes sequentially from the input atmospheric parameters through a series of intermediate layers to the predicted spectrum, with no recurrent connections between layers. The four atmospheric parameters are provided as inputs to the network: $T_{\mathrm{eff}}$, $\log g$, $[\mathrm{Fe}/\mathrm{H}]$, and $\xi$. The network contains two intermediate (or ''hidden'') layers of 256 computational units (neurons) each. Each neuron forms a weighted combination of the outputs from the preceding layer and applies a rectified linear unit (ReLU) activation function, $\max(0,x)$, allowing the network to represent nonlinear variations of the spectrum with stellar parameters. The final layer contains one output for each wavelength pixel within the selected Fe line windows, so that the complete set of line profiles is predicted simultaneously for a given set of atmospheric parameters.

The network is trained on line depths, $1-F_\lambda$, rather than directly on normalized flux. Before training, each atmospheric parameter is standardized by subtracting its mean and dividing by its standard deviation across the synthetic grid. The line depths are standardized similarly, using a separate mean and standard deviation at each wavelength pixel. These transformations place the inputs and outputs on comparable numerical scales and improve the stability of the optimization.

Training minimizes a weighted mean-squared-error loss between the emulator predictions and the synthetic training spectra, with larger weights assigned near the Fe line cores so that the pixels most sensitive to iron abundance contribute more strongly to the optimization. The network parameters are optimized with \texttt{AdamW} \citep{adamw}. One training epoch corresponds to one complete pass through the training set, and training is allowed to proceed for up to 200 epochs. To limit overfitting, optimization is stopped if the loss measured on the independent validation set does not improve for 20 consecutive epochs, and the network parameters corresponding to the minimum validation loss are retained.

The accuracy of the resulting emulator is quantified from its reconstruction residuals on the held-out validation spectra. At each wavelength pixel, we calculate the difference between the emulator prediction and the corresponding \texttt{Turbospectrum} flux for every validation spectrum. The resulting per-pixel reconstruction uncertainty, $\sigma_{\mathrm{recon}}$, therefore measures the typical error introduced by replacing direct spectral synthesis with the neural network prediction. We estimate this scatter robustly using the median absolute deviation (MAD) of the validation residuals and multiply it by $1.4826$. This factor converts the MAD to the equivalent standard deviation for a Gaussian distribution. The resulting $\sigma_{\mathrm{recon}}$ is propagated in the spectroscopic likelihood so that interpolation errors from the emulator are included in the model uncertainty.

During the stellar parameter fit, flat priors on $[\mathrm{Fe}/\mathrm{H}]$ and $v\sin i$ are combined with a likelihood based on the differences between the observed and model fluxes to define their joint posterior probability distribution. For each trial pair of $[\mathrm{Fe}/\mathrm{H}]$ and $v\sin i$, the emulator predicts the spectrum at the trial metallicity and the fixed photometric values of $T_{\mathrm{eff}}$, $\log g$, and $\xi$. The predicted spectrum is then rotationally broadened according to the trial $v\sin i$ and compared with the observed spectrum over all accepted Fe line windows.

For each trial value of $v\sin i$, the emulator-predicted flux spectrum is rotationally broadened by convolution with a linear limb-darkened kernel \citep{Graybook},
\begin{eqnarray} K(\Delta v) = \frac{ 2(1-\epsilon)\sqrt{1-[\Delta v/(v\sin i)]^2} } {\pi (v\sin i)(1-\epsilon/3)} \nonumber\\ + \frac{ \frac{\pi\epsilon}{2}\left[1-[\Delta v/(v\sin i)]^2\right] } {\pi (v\sin i)(1-\epsilon/3)} . \end{eqnarray}
Here, $\epsilon=0.6$ is the adopted limb-darkening coefficient, $\Delta v$ is the velocity offset within the convolution kernel, and $v\sin i$ is the projected rotational velocity. The expression applies for $|\Delta v|\leq v\sin i$, while $K(\Delta v)=0$ otherwise. The kernel is normalized to unit area before convolution.

Observed spectra are locally pseudo-continuum normalized using synthetic predictions as a guide. Within each local line window, pixels with synthetic flux $F_{\mathrm{syn}}>0.999$ and $|\partial F_{\mathrm{syn}}/\partial\lambda|<0.002$ are flagged as candidate continuum points. Assuming that the pseudo-continuum varies approximately linearly across these narrow wavelength intervals, a single $2\sigma$ clipping step is applied to these points before fitting a first-order polynomial. This polynomial is then used to divide both the observed fluxes and their uncertainties. For each Fe line, the fitting window is defined from a reference synthetic spectrum generated at the current stellar parameters. The line core is identified as the minimum synthetic flux within the allowed line region, and the line edges are taken as the nearest points on either side where $|\partial F_{\mathrm{syn}}/\partial\lambda|<0.002$.

Before jointly sampling the posterior distribution of $[\mathrm{Fe}/\mathrm{H}]$ and $v\sin i$, each line is fit on a coarse $[\mathrm{Fe}/\mathrm{H}]$--$v\sin i$ grid to assess its behavior. Lines are rejected if (i) they contain fewer than five pixels in the final fitting window, (ii) the best-fit $[\mathrm{Fe}/\mathrm{H}]$ lies within 0.05 dex of a grid edge, (iii) the minimum reduced $\chi^2$ exceeds a permissive quality control threshold defined as 25 times the expected reduced-$\chi^2$ scatter, $\sigma(\chi_\nu^2)=\sqrt{2/\nu}$, where $\nu$ is the number of degrees of freedom, (iv) the normalized posterior Shannon entropy, $H=-\sum_i p_i\ln p_i/\ln N$, where $p_i$ is the normalized posterior probability at the $i$th grid point and $N$ is the total number of grid points, exceeds 0.90; the quantity $H$ measures how broadly the posterior probability is distributed, with values approaching unity indicating a nearly flat posterior, or (v) the $\chi^2$ profile is monotonic and therefore has no well-defined minimum. 

Only lines that pass these quality-control criteria are included in the joint fit, in which all accepted Fe lines simultaneously constrain $[\mathrm{Fe}/\mathrm{H}]$ and $v\sin i$. The likelihood, up to an additive constant, is defined by summing the pixel-level residuals over all accepted lines:
\begin{equation}
\ln \mathcal{L}([\mathrm{Fe}/\mathrm{H}], v\sin i)
= -\frac{1}{2} \sum_{\ell=1}^{N_{\rm lines}}
\sum_{i=1}^{N_\ell}
\frac{\bigl(F_{\mathrm{obs},i,\ell} - F_{\mathrm{syn},i,\ell}\bigr)^2}
{\sigma_{\mathrm{obs},i,\ell}^2 + \sigma_{\mathrm{recon},i,\ell}^2} ,
\end{equation}
where the outer sum runs over the $N_{\rm lines}$ accepted Fe lines and $N_\ell$ is the number of pixels in the fitting window of line $\ell$. The quantities $F_{\mathrm{obs},i,\ell}$ and $F_{\mathrm{syn},i,\ell}$ are the observed and synthetic normalized fluxes, respectively, while $\sigma_{\mathrm{obs},i,\ell}$ and $\sigma_{\mathrm{recon},i,\ell}$ are the observational and emulator reconstruction uncertainties, respectively. The reconstruction uncertainties are interpolated onto the observed wavelength grid before evaluating the likelihood.

The resulting posterior distribution is sampled with \texttt{emcee} \citep{emcee} using flat priors, $-2.5 \leq [\mathrm{Fe}/\mathrm{H}] \leq 0.5$ and $0 \leq v\sin i \leq 50~\mathrm{km\,s^{-1}}$. The final $[\mathrm{Fe}/\mathrm{H}]$ and $v\sin i$ values are taken as the medians of their marginalized posterior distributions, with uncertainties derived from the 16th--84th percentile ranges.

The iterative determination of the stellar parameters alternates between the photometric and spectroscopic steps described above. In the first cycle, the photometric fit is initialized with solar metallicity and uncertainity 0.2 dex, adopted as a broad initial prior to avoid strongly constraining the solution; the resulting $T_{\rm eff}$, $\log g$, and $\xi$ are then fixed while [Fe/H] and $v \sin i$ are determined spectroscopically. In subsequent cycles, the spectroscopic [Fe/H] is used as the metallicity in the next photometric fit. Iterations continue until the difference between spectroscopic and photometric [Fe/H] is smaller than the photometric metallicity uncertainty, typically requiring 2–5 iterations per star. Diagnostic figures are generated after every iteration to validate convergence and identify problematic lines.

Derived atmospheric parameters ($T_{\rm eff}$, $\log g$, $[\mathrm{Fe/H}]$, $\xi$, and $v\sin i$) and their uncertainties are listed in Table~\ref{table::params_abuns2}. 
The reported $[\mathrm{Fe/H}]$ are referenced to the solar abundance scale of \citet{Asplund2009}. The distribution of stars in the Kiel diagram, color-coded by metallicity, is shown in Figure~\ref{fig:kiel}, illustrating that our sample spans both dwarf and giant evolutionary stages. The ranges of stellar parameters are $T_{\rm eff} \sim 3800$--$6800$~K, $\log g \sim 1.0$--$4.8$~dex, and $[\mathrm{Fe/H}] \sim -1.5$ to $0.5$. Typical internal $1\sigma$ uncertainties are $\sim$10--30~K in $T_{\rm eff}$, $\sim$0.01--0.05~dex in $\log g$, and $\sim$0.01--0.03~dex in $[\mathrm{Fe/H}]$ from the spectral analysis.

\begin{deluxetable*}{ c c c c c c c c c c c}
\tablewidth{0pt} 
\tabletypesize{\footnotesize}
\tablecaption{Atmospheric Parameters and P Abundances \label{table::params_abuns2}} 
\tablehead{
\colhead{Name} & \colhead{Teff} & \colhead{$\sigma$ Teff} & \colhead{log(g)} & \colhead{$\sigma$ log(g)} & \colhead{[Fe/H]} & \colhead{$\sigma$ [Fe/H]} & \colhead{$\xi$} & \colhead{vsini} & \colhead{[P/Fe]} & \colhead{$\sigma$ [P/Fe]} \\
\colhead{} & \colhead{(K)} & \colhead{(K)} & \colhead{(dex)} & \colhead{(dex)} & \colhead{(dex)} & \colhead{(dex)} & \colhead{(km s${^{-1}}$)} & \colhead{(km s${^{-1}}$)} & \colhead{(dex)} & \colhead{(dex)}
}
\startdata
2MASS J00230934+0007429 & 5033 & 13 & 3.37 & 0.03 & -0.03 & 0.01 & 1.13 & 2.96 & -0.161 & 0.076 \\
2MASS J01062733+5729443 & 4955 & 10 & 2.88 & 0.02 & -0.31 & 0.01 & 1.37 & 8.99 & $<$-0.441 & - \\
2MASS J03065740+5449428 & 4584 & 11 & 2.33 & 0.03 & -0.42 & 0.04 & 1.55 & 5.46 & 0.266 & 0.162 \\
2MASS J03221188+2736271 & 4860 & 19 & 2.17 & 0.06 & -0.01 & 0.01 & 1.59 & 9.37 & -0.006 & 0.033 \\
2MASS J04383875-0258161 & 4889 & 16 & 2.55 & 0.03 & -0.10 & 0.03 & 1.49 & 3.58 & -0.096 & 0.086 \\
$\cdots$ & $\cdots$ & $\cdots$ & $\cdots$ & $\cdots$ & $\cdots$ &
$\cdots$ & $\cdots$ & $\cdots$ & $\cdots$ & $\cdots$ \\
\enddata
\tablecomments{(This table is available in its entirety in machine-readable form.)}
\end{deluxetable*}

\subsubsection{Atmospheric parameters validation} \label{sec:sec3.1.1}

\begin{figure*}
    \centering
\includegraphics[width=\textwidth]{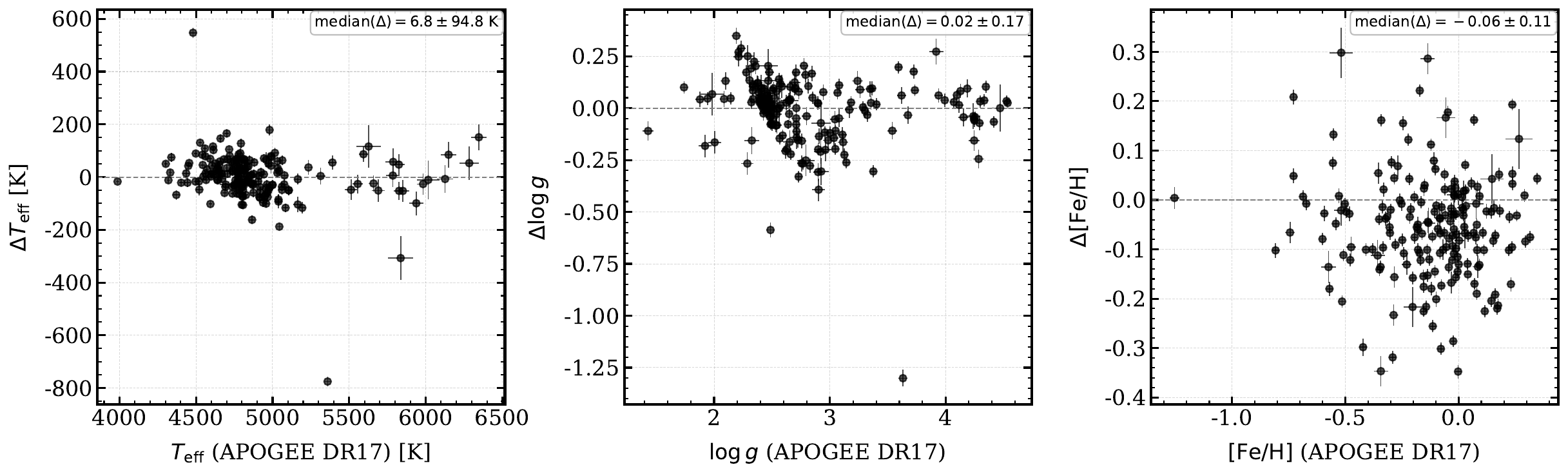}
    \caption{Residuals between the atmospheric parameters derived in this work and the corresponding APOGEE DR17 measurements are shown for stars in our sample with available APOGEE data. The residuals are defined as the value from this work minus the APOGEE DR17 value and are plotted against the APOGEE DR17 parameters, with $T_{\rm eff}$, $\log g$, and $[\mathrm{Fe/H}]$ shown from left to right. Error bars represent the combined uncertainties from this work and APOGEE DR17 added in quadrature. The annotation indicates the median offset and the standard deviation of the scatter.}
    \label{fig:comparison_apogee}
\end{figure*}

\begin{figure*}
    \centering
    \includegraphics[width=\textwidth]{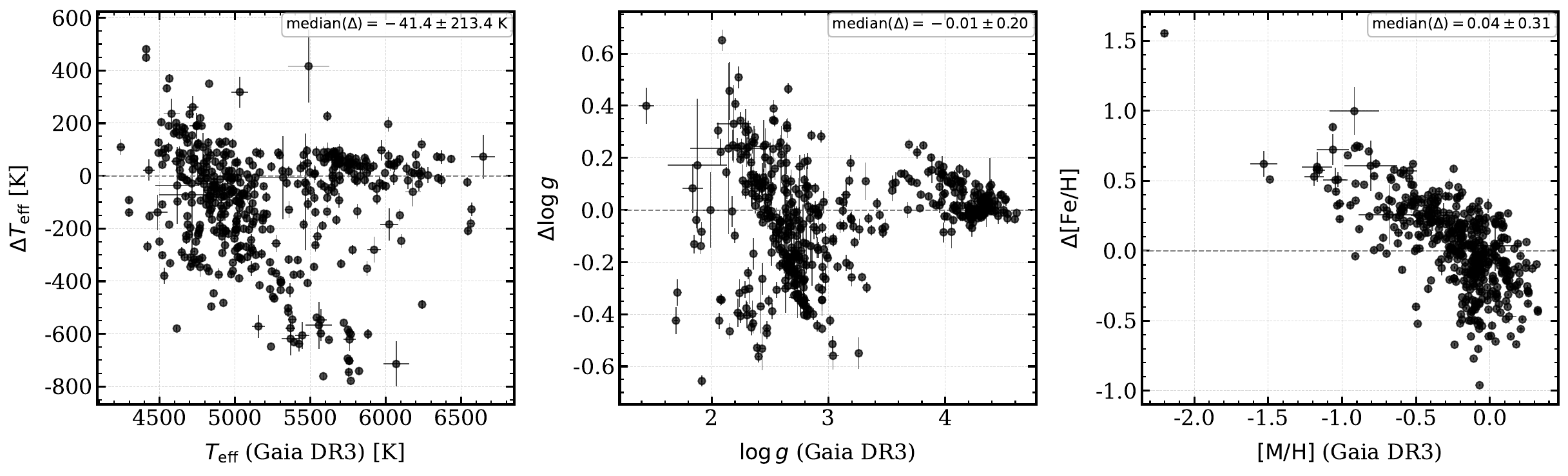}
    \caption{Residuals between the atmospheric parameters derived in this work and the corresponding Gaia DR3 GSP-Phot measurements are shown for the subset of stars in our sample with available Gaia data. The residuals are defined as the value from this work minus the Gaia DR3 value and are plotted against the Gaia DR3 parameters, with $T_{\rm eff}$, $\log g$, and $[\mathrm{Fe/H}]$ shown from left to right. Error bars represent the combined uncertainties from this work and Gaia added in quadrature. The annotation indicates the median offset and the standard deviation of the scatter.}
    \label{fig:comparison_gaia}
\end{figure*}

As an external validation of our adopted stellar parameters, we performed two complementary consistency checks against literature determinations. First, we performed comparisons against large homogeneous surveys that employ internally consistent analysis frameworks. For stars in our sample with available measurements, we compared our results to APOGEE DR17 \citep{apogeedr17} and Gaia DR3 \citep{GaiaDr3} parameters. For Gaia DR3, we used the GSP-Phot estimates of $T_{\mathrm{eff}}$, $\log g$, and $[\mathrm{Fe}/\mathrm{H}]$. These subsamples are necessarily smaller, as not all stars have corresponding measurements in these surveys. The comparison with APOGEE, based on 199 stars (Figure~\ref{fig:comparison_apogee}), shows reduced scatter, with median residuals of $\Delta T_{\rm eff} \sim 7 \pm 95$~K, $\Delta \log g \sim 0.02 \pm 0.17$, and $\Delta[\mathrm{Fe/H}] \sim -0.06 \pm 0.11$~dex. The median APOGEE uncertainties for these stars are 9.3~K in $T_{\rm eff}$, 0.024 in $\log g$, and 0.008~dex in [Fe/H]. Since the observed residual dispersions are larger than expected from the formal uncertainties of either analysis, the quoted internal uncertainties are likely underestimated to some extent. Similarly, the Gaia DR3 comparison, based on 467 stars (Figure~\ref{fig:comparison_gaia}), yields $\Delta T_{\rm eff} \sim -41 \pm 213$~K, $\Delta \log g \sim -0.01 \pm 0.20$, and $\Delta[\mathrm{Fe/H}] \sim 0.04 \pm 0.31$~dex. While Gaia shows larger dispersion—particularly in effective temperature—this is consistent with the lower spectral resolution and broader methodological differences relative to high-resolution spectroscopy. Overall, these comparisons demonstrate good agreement with independent, homogeneous analyses and confirm the robustness of our parameter determination.

Discrepancies between photometric and spectroscopic stellar parameters are expected and reflect both physical and methodological limitations. Spectroscopic determinations typically rely on excitation and ionization balance under LTE assumptions, which can introduce systematic biases, particularly in metal-poor stars (e.g., \citealt{Cayrel2004,Frebel2013,Ezzeddine2020}). Photometric estimates, while largely independent of these assumptions, depend on empirical calibrations and accurate extinction corrections. The overall level of agreement across all comparisons, despite these differing approaches, supports the reliability of the adopted atmospheric parameters.

\begin{figure*}
    \centering
    \includegraphics[width=\textwidth]{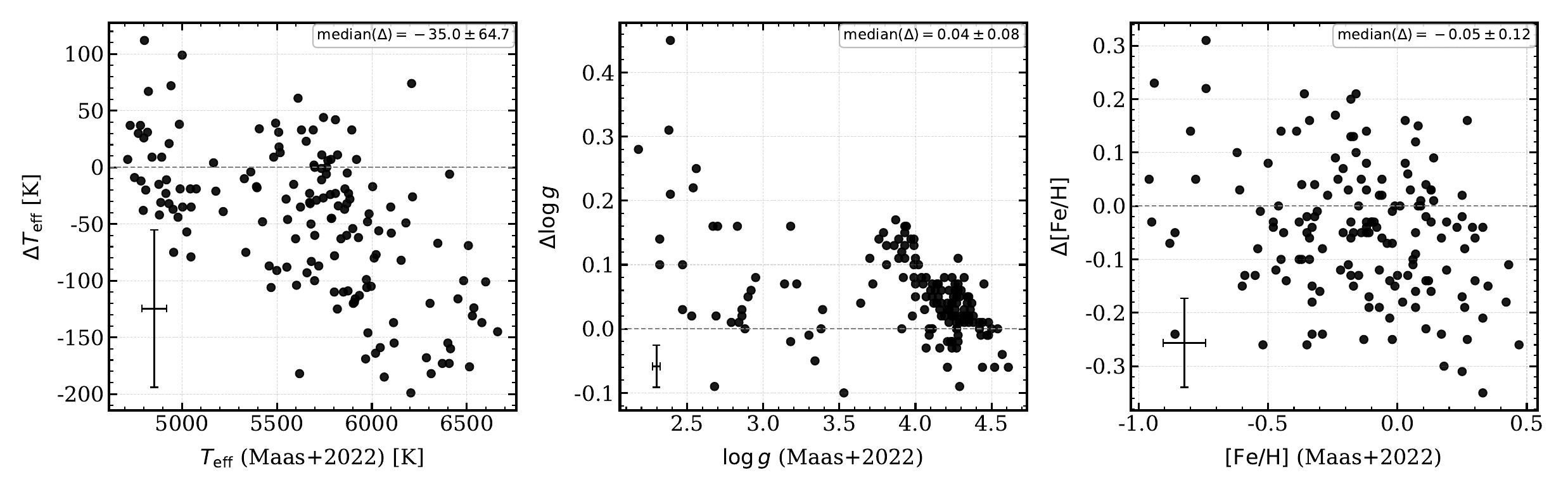}
    \caption{Residuals between the atmospheric parameters derived in this work and those from \citet{Maas2022} for the 154 stars in common. The residuals are defined as the value from this work minus the \citet{Maas2022} value and are plotted against the corresponding \citet{Maas2022} parameters, with $T_{\rm eff}$, $\log g$, and $[\mathrm{Fe/H}]$ shown from left to right. The error cross in the lower-left corner of each panel shows the mean \citet{Maas2022} uncertainty along the horizontal axis and the mean residual uncertainty along the vertical axis, with the latter obtained by adding the uncertainties from this work and \citet{Maas2022} in quadrature. The annotation gives the median offset and the standard deviation of the residuals.
    }
    \label{fig:comparison_maas22_params}
\end{figure*}

As a second test, we used the sample of \citet{Maas2022} and compared our derived stellar parameters to those reported in that study, which employs a similar spectro-photometric approach but differs in inference methodology and line analysis. \citet{Maas2022} employed the \texttt{MINESweeper} framework \citep{Cargile2020minesweeper} for stellar parameter inference and \texttt{MOOG}-based \citep{moog} equivalent width measurements of Fe lines. The residuals are shown in Figure~\ref{fig:comparison_maas22_params}. For effective temperature, we find a small systematic offset, with a median $\Delta T_{\rm eff}$ of $-35 \pm 65$ K. Our temperatures tend to be slightly cooler than those of \citet{Maas2022}, particularly toward the hotter (mainly dwarf) end of the sample, whereas the agreement is tighter among cooler giants. For surface gravity, the median offset is small, $\Delta \log g = 0.04 \pm 0.08$ dex. The scatter is larger among giants (lower $\log g$), where our derived $\log g$ are systematically higher relative to \citet{Maas2022}, while the agreement for dwarfs is notably tighter and largely centered around zero. This behavior likely reflects the increased degeneracy between $T_{\rm eff}$ and $\log g$ in evolved stars. We find excellent overall consistency for metallicities, with a median $\Delta$[Fe/H] of $-0.05 \pm 0.12$ dex and no strong systematic trend across the metallicity range. 

\subsection{Abundances derivation methodology}  \label{sec:sec3.2}

Elemental abundances were determined by fitting synthetic spectra to selected absorption features through a $\chi^2$ minimization procedure, with results expressed as abundance ratios relative to iron, $[X/{\rm Fe}]$. We adopt spectral synthesis rather than equivalent width methods, as the main line of interest (P I) is intrinsically weak and often can be blended \citep{Maas2022}, rendering equivalent widths less reliable \citep{Jofre2019}.

For each trial abundance ratio $x_j \equiv [X/{\rm Fe}]_j$, where $X$ denotes an element other than Fe, sampled over a grid (typically $-0.9 < x_j < +0.9$ in steps of 0.3 dex), a synthetic spectrum is computed with \texttt{Turbospectrum} \citep{turbospectrum}. The synthetic calculations adopt the final $T_{\rm eff}$, $\log g$, [Fe/H], and $\xi$ values derived in Section~\ref{sec:sec3.1}. The atmospheric structure is interpolated from the MARCS grid \citep{Gustafsson2008}, adopting plane-parallel geometry for high-gravity stars and spherical geometry for low-gravity stars \citep{Masseron2006}. 
Prior to the $\chi^2$ computation, the observed spectra were pseudo-continuum normalized with the procedure outlined in Section~\ref{sec:sec3.1}.

A dynamically defined fitting window is used, determined from the synthetic spectrum's flux gradient to isolate the target line from nearby blends and continuum regions. Within this window, we first compute $\chi^2$ over the grid of trial abundance ratios $x_j$:
\begin{equation}
\chi^2(x_j) = \sum_{i \in {\rm window}}
\left(
\frac{F_{{\rm obs},i}-F_{{\rm syn},i}(x_j)}
{\sigma_{{\rm obs},i}}
\right)^2 .
\end{equation}
Here, $j$ indexes the trial abundance ratio $x_j$ on the coarse grid; $i$ indexes the pixels within the dynamically defined fitting window; $F_{{\rm obs},i}$ is the observed normalized flux in the $i$th pixel; $F_{{\rm syn},i}(x_j)$ is the synthetic normalized flux evaluated at the wavelength of the $i$th pixel for the trial abundance ratio $x_j$; and $\sigma_{{\rm obs},i}$ is the observational flux uncertainty in the $i$th pixel. The $\chi^2$ value is computed using all pixels within the fitting window for each trial abundance ratio, and the value that minimizes $\chi^2(x_j)$ is adopted as the best-fit abundance ratio, $x_{\rm best}$.

In some cases the line fit does not provide a robust measurement. When the fit diagnostics indicate that the line is either not significantly detected or does not yield a well-constrained abundance, we instead report an upper limit. This occurs when: (i) the best-fit abundance ratio lies at the grid edge; (ii) the Markov chain Monte Carlo (MCMC)-derived posterior distribution (described below) is broad or flat (the 68\% credible interval wider than 0.6 dex); (iii) the $\chi^2$ curve is monotonic; (iv) the best-fit $\chi^2/N_{pix}$ exceeds 5 in the fitting window, where $N_{pix}$ is the number of pixels in the fitting window; or (v) the normalized posterior Shannon entropy exceeds 0.90 (indicating flatness). We define the 95\% confidence threshold by

\begin{equation}
\chi^2(x_j) - \chi^2_{\min} = 3.84,
\end{equation}
and take the upper limit $x_{\rm limit}$ as the largest abundance ratio on the sampled grid that satisfies this criterion (with $x_{\rm limit} > x_{\rm best}$).

Uncertainties in $[X/{\rm Fe}]$ were computed as the quadrature sum of two components. The first one is the parameter-sensitivity uncertainty, $\sigma_{\rm sens}$, which reflects the propagated effect of uncertainties in the adopted stellar parameters. It is evaluated by perturbing each parameter $p$ in the set $\{T_{\rm eff},\log g,[{\rm Fe}/{\rm H}],\xi\}$ by its corresponding $\pm1\sigma_p$ uncertainty, while holding the remaining parameters fixed, and re-deriving the abundance ratio $x\equiv[X/{\rm Fe}]$. The resulting change in $x$ quantifies the sensitivity of the abundance measurement to the uncertainty in that parameter. The contributions are combined as
\begin{equation}
\sigma_{\rm sens}^2 = \sum_{p} \left( \frac{1}{2}  |x(p+\sigma_p) - x(p-\sigma_p)| \right)^2 .
\end{equation}
Here, $x(p+\sigma_p)$ and $x(p-\sigma_p)$ are the abundance ratios re-derived after increasing or decreasing each parameter $p$ by its $1\sigma$ uncertainty, with all other parameters held fixed. The second uncertainty component, statistical fitting uncertainty, $\sigma_{\rm fit}$, quantifies the precision of the abundance measurement for fixed stellar parameters. It is derived by sampling the posterior distribution
\begin{equation}
\ln \mathcal{L}(x) = -\frac{1}{2} \chi^2(x)
\end{equation}
with a flat prior using MCMC, and taking half the difference between the 84th and 16th percentiles. The total uncertainty is then
\begin{equation}
\sigma_{\rm total} = \sqrt{\sigma_{\rm sens}^2 + \sigma_{\rm fit}^2}.
\end{equation}

For each line, the final reported abundance is $[X/{\rm Fe}] = x_{\rm best} \pm \sigma_{\rm total}$ when the fit passes all acceptance criteria, or an upper limit $[X/{\rm Fe}] < x_{\rm limit}$ otherwise. All abundances are reported relative to the solar reference scale of \citet{Asplund2009}.

\subsection{P abundances}  \label{sec:sec3.3}

The most prominent P~I feature in the Y-band is located at 10529.52~\AA\ and serves as the diagnostic for P abundance determinations in this work. The atomic data adopted for this transition are listed in Table~\ref{table::linelist}. Previous work has demonstrated the ability of this line to yield precise P abundances \citep{Sneden2021,Maas2022}. The feature lies in a relatively clean telluric region and is well separated from nearby CN and Ni~I lines, as confirmed through inspection of the Arcturus atlas \citep{Hinkle1995}. Other candidate P~I lines were excluded because they fall within spectral gaps, are intrinsically weaker at the signal-to-noise ratios of our data, or suffer from significant telluric blending. The derived P abundances and their associated uncertainties (or upper limits) for our stars are reported in Table~\ref{table::params_abuns2}. 
All spectra were inspected visually to verify the quality of the continuum placement and the reliability of the spectral fits.

\begin{figure*}
    \centering
    \includegraphics[width=0.8\textwidth]{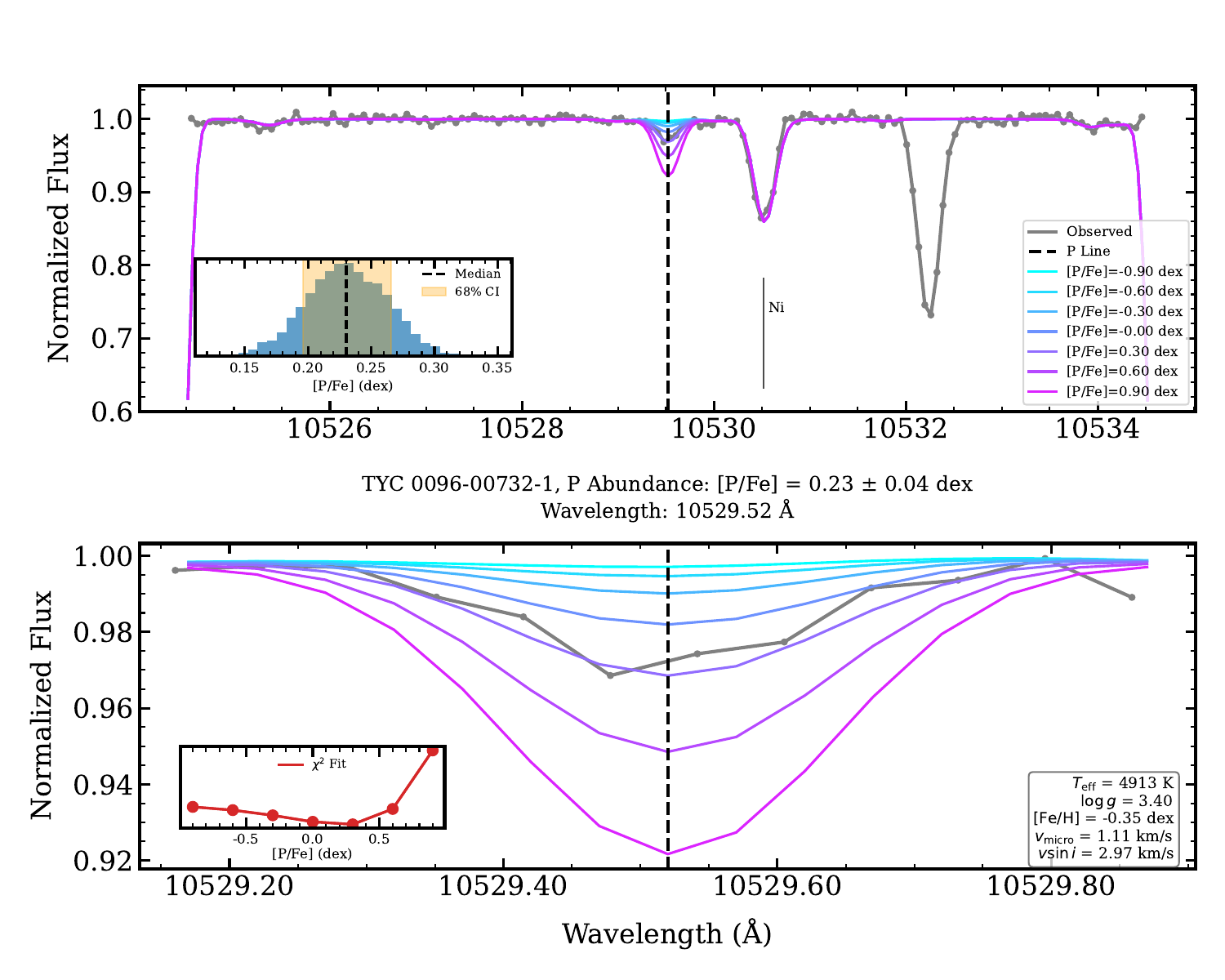}
    \caption{
    Example P abundance determination from the P~I 10529.52~\AA\ line for the star TYC 0096-00732-1. The top panel shows the observed spectrum (gray) together with synthetic spectra computed for a range of [P/Fe] values (colored curves, spaced by 0.30 dex). The vertical dashed line marks the P~I line center, while nearby features are also indicated (e.g., Ni). The inset in the top panel displays the posterior distribution from the MCMC fit, with the median and 68\% credible interval indicated. The bottom panel shows a zoomed-in view of the fitting region around the P line, while its inset shows the corresponding $\chi^2$ as a function of [P/Fe]. The stellar parameters adopted for the fit are listed in the lower-right corner.
}
    \label{fig:example_fit}
\end{figure*}

An example P abundance determination from the P I 10529.52~\AA\ line is shown in Figure~\ref{fig:example_fit} for the star TYC 0096-00732-1, illustrating the sensitivity of the line depth (typically a few percent) to abundance variations. The top panel displays the observed spectrum (gray) together with synthetic spectra computed for a range of [P/Fe] values (colored curves, spaced by 0.30 dex) at the star's derived atmospheric parameters. The vertical dashed line marks the P I line center, and nearby blending features (e.g., Ni) are indicated. The bottom panel shows a zoomed-in view of the fitting region around the P feature. 
The inset in the top panel presents the posterior distribution from the MCMC fit, with the median value and 68\% credible interval indicated, while the inset in the bottom panel shows the corresponding $\chi^2$ as a function of [P/Fe]. The adopted stellar parameters are listed in the lower-right corner. We note that the Fe~I feature near 10532.11~\AA\ is not reproduced accurately by the synthetic spectra, likely due to uncertainties in the adopted oscillator strengths in the line list. However, this discrepancy occurs sufficiently far from the P I line and does not affect the abundance determination.

\subsubsection{P abundances validation} \label{sec:sec3.2.1}

\begin{figure}
    \centering
\includegraphics[width=0.45\textwidth]{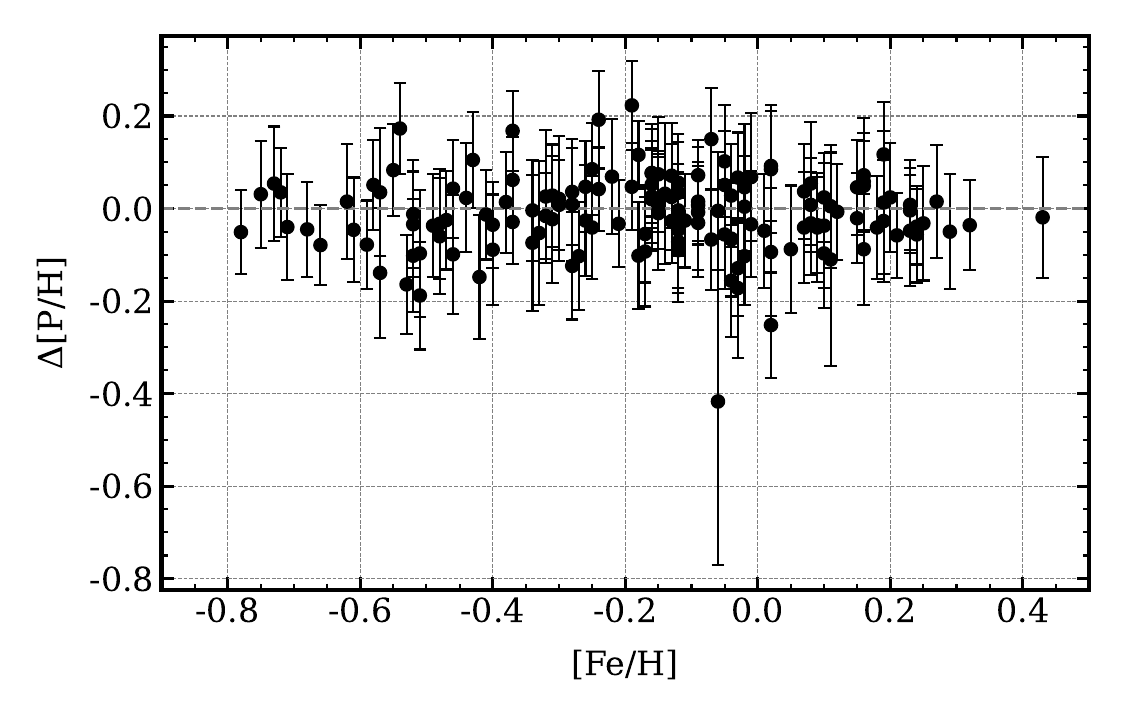}
    \caption{
Comparison of P abundances derived in this work with those reported by \citet{Maas2022} for the same 154 stars. Shown are residuals $\Delta[\mathrm{P/H}]$ (this work minus \citealp{Maas2022}) as a function of [Fe/H]. Error bars reflect the quadrature sum of the individual abundance uncertainties from both studies. The median residual is $-0.032 \pm 0.084$ dex, with no systematic trend across the metallicity range.
}
    \label{fig:P_maas22_comparison1}
\end{figure}

We applied our abundance analysis to the same sample of 154 stars previously analyzed by \citet{Maas2022}, all observed with HPF. Adopting our derived stellar parameters, P abundances were re-determined exclusively from the P I 10529.52~\AA\ line following the procedure described in Section~\ref{sec:sec3.2}. The resulting abundances were compared directly to those reported by \citet{Maas2022}. Figure~\ref{fig:P_maas22_comparison1} shows the residuals $\Delta[\mathrm{P/H}]$ as a function of metallicity. The \citet{Maas2022} P abundances were homogenized by rescaling their adopted solar P abundance to the \citet{Asplund2009} scale used in this work. 
We find close agreement between the two studies, with a median offset of $-0.012 \pm 0.084$ dex and no systematic trend with [Fe/H]. For comparison, the typical [P/Fe] uncertainties are $0.037$ dex in this work and $0.067$ dex in \citet{Maas2022}. 
The scatter is readily explained by differences in radiative-transfer implementation, adopted 
linelists, continuum normalization (particularly important for a weak line such as P I 10529.52~\AA), and fitting methodology \citep{Jofre2015,Hinkel2016}. As an additional test, we repeated the analysis adopting the stellar parameters of \citet{Maas2022} and obtained statistically indistinguishable residuals, further confirming the consistency of our analysis.


\begin{figure}
    \centering
\includegraphics[width=0.45\textwidth]{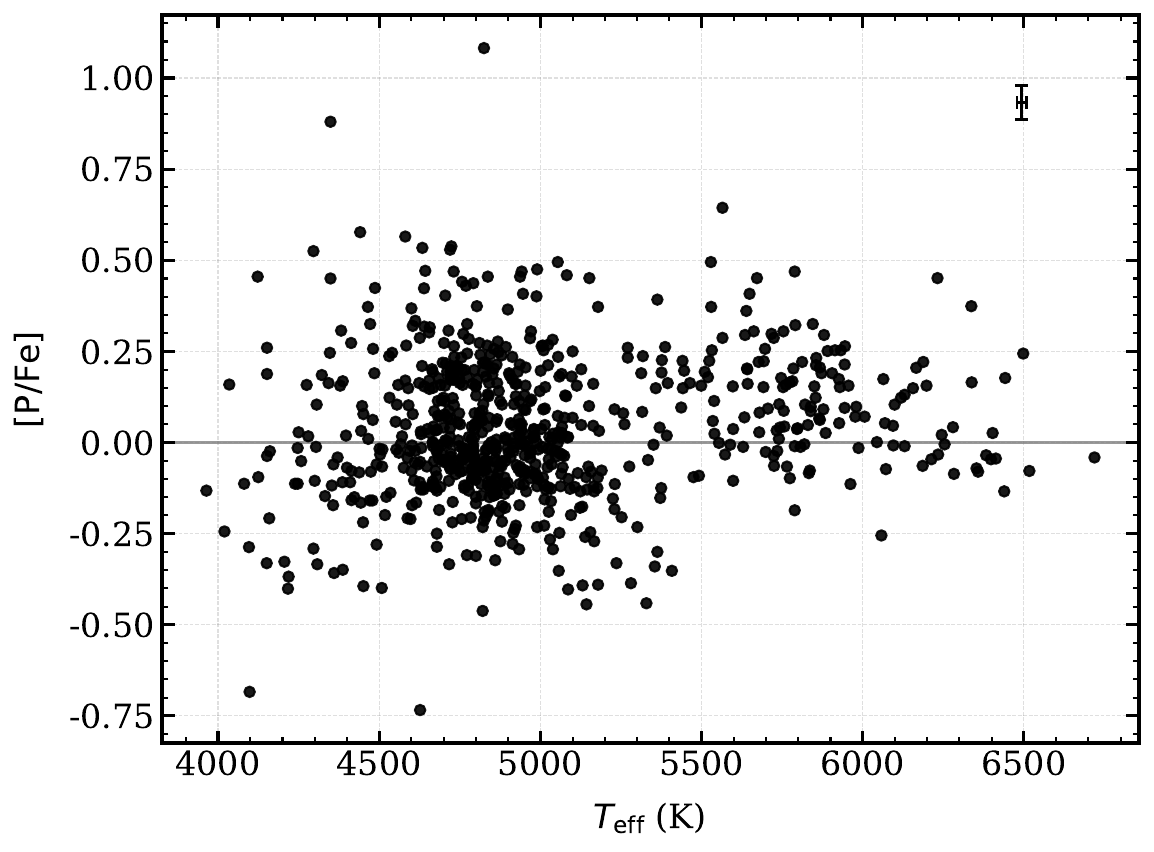}
    \caption{
P abundances as a function of effective temperature for stars in our sample. The mean uncertainties in $T_{\rm eff}$ and $[\mathrm{P}/\mathrm{Fe}]$ are shown in the upper-right corner. The absence of a clear trend with $T_{\rm eff}$ suggests that temperature-dependent systematics are not a significant source of bias in the abundance determinations across the parameter range considered.
}
\label{fig:pfe_teff}
\end{figure}

Finally, we find only a weak positive trend between [P/Fe] and effective temperature (Figure~\ref{fig:pfe_teff}), with a linear slope of +0.052 dex per 1000 K. This trend may partly reflect the underlying distribution of stellar parameters in our sample, rather than a strong temperature-dependent systematic in the inferred P abundances. The absence of a pronounced trend with effective temperature therefore suggests that unaccounted non-LTE effects are unlikely to be the dominant driver of the observed abundance behavior, in agreement with earlier analyses of this line \citep{Maas2022}. While non-LTE effects may in principle affect P I lines, such corrections have only recently become available \citep{Andrievsky2026}. To assess the impact of non-LTE effects, we evaluated the corrections of \citet{Andrievsky2026} for the P I 10529.52~\AA\ transition using four-dimensional linear interpolation in effective temperature, surface gravity, metallicity, and [P/Fe]. The corrections result in a systematic downward shift of the derived abundances by $\sim 0.15$--$0.2$~dex, while 
preserving the overall [P/Fe]--[Fe/H] trends and the relative behavior of the stellar populations. Given our focus on differential comparisons between Galactic components and the characterization of global abundance trends, we retain LTE abundances throughout this study, consistent with most previous P abundance studies.


\section{Results}  \label{sec:sec4}

\subsection{[P/Fe] general trend} \label{sec:sec4.1}

\begin{figure*}
    \centering
    \includegraphics[width=0.99\textwidth]{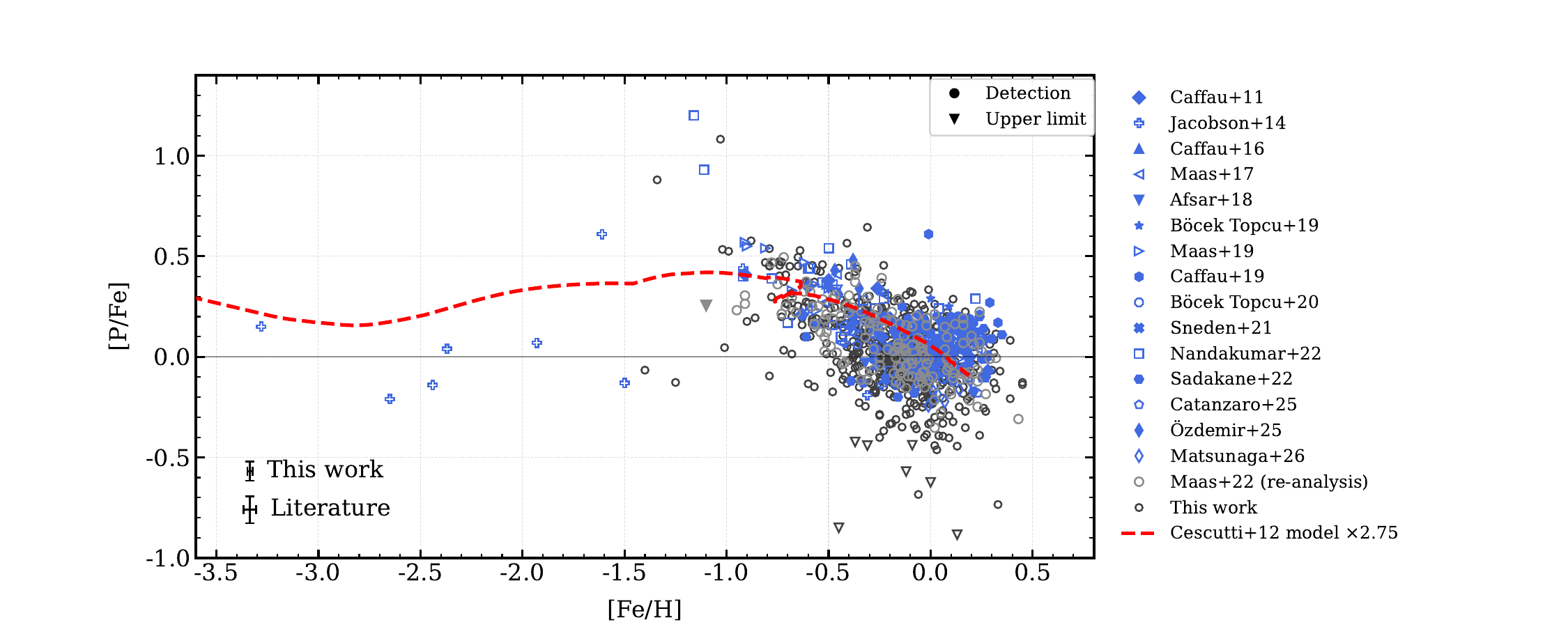}
    \caption{P abundances as a function of metallicity for our sample compared to literature measurements. Literature data from multiple studies are shown in blue, while the re-analyzed \citet{Maas2022} sample is shown in gray, and results from this work are shown in black. For the abundances derived in this study, filled circles denote abundance detections, while inverted triangles indicate upper limits. The scaled Galactic chemical evolution model of \citet{Cescutti2012}, multiplied by a factor of 2.75 in [P/Fe], is shown as a dashed red curve. The distribution of literature sources is indicated in the legend. Representative median uncertainties for this work and the literature are shown in the lower-left corner.}
    \label{fig:pfe}
\end{figure*}

\begin{figure*}
    \centering
    \includegraphics[width=0.95\textwidth]{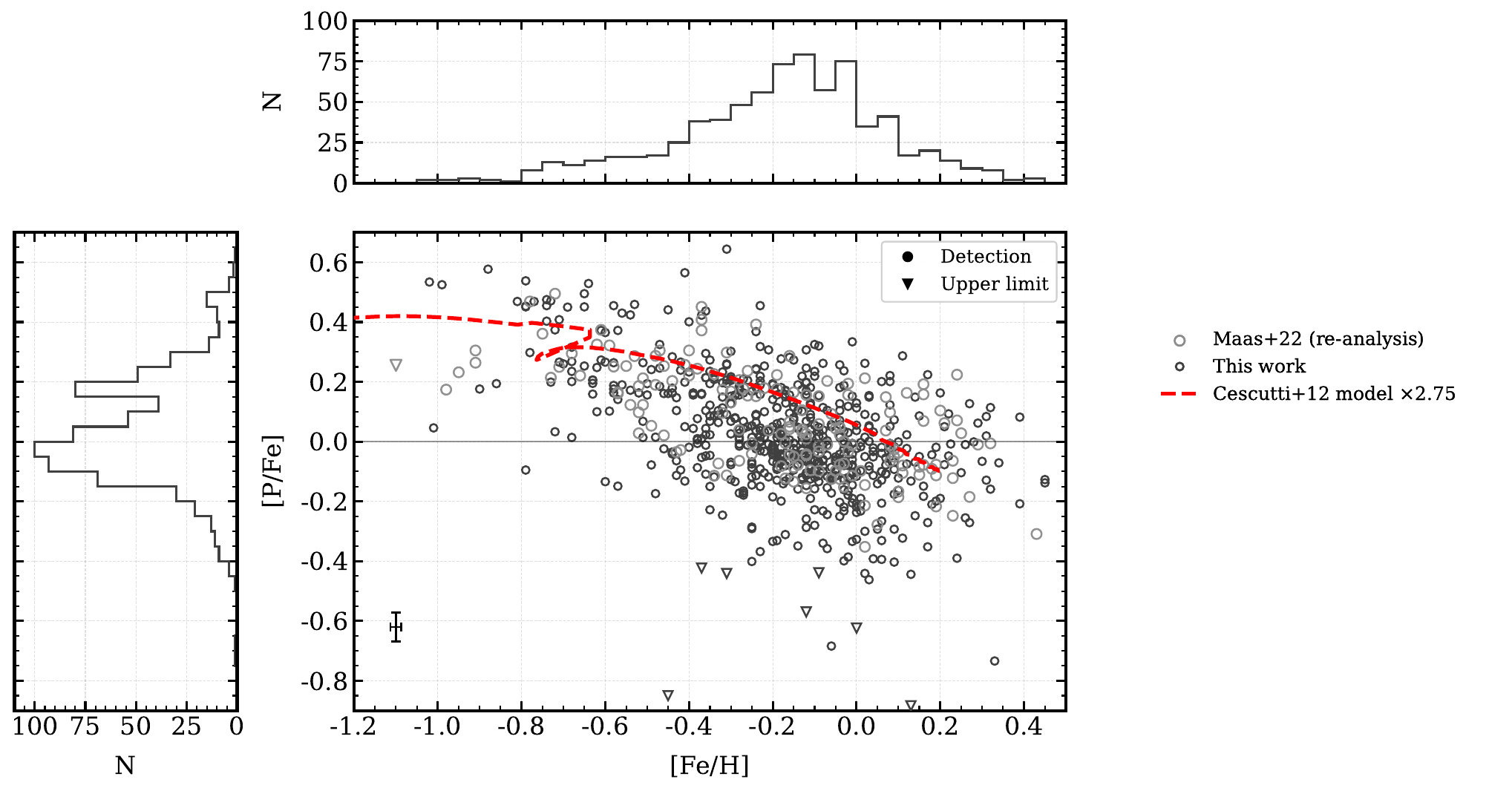}
    \caption{Enlarged view of Figure~\ref{fig:pfe}, focusing on the metallicity and abundance range containing the majority of the stars analyzed in this work. Only measurements from this work and the re-analysis of the \citet{Maas2022} sample are shown. The central panel displays [P/Fe] as a function of [Fe/H], with circles indicating detections and inverted triangles indicating upper limits. The marginal histograms show the [Fe/H] and [P/Fe] distributions for detections from this work only. The scaled Galactic chemical evolution model of \citet{Cescutti2012}, with the P yields multiplied by a factor of 2.75, is shown as a dashed red curve. The median uncertainties are shown in the lower-left corner.}
    \label{fig:pfe_zoom}
\end{figure*}

P abundances as a function of metallicity are shown in Figure~\ref{fig:pfe}, where we compare our results with literature measurements, including the re-analysis of the \citet{Maas2022} sample. The median uncertainties for our sample are $0.013$ dex in [Fe/H] and $0.048$ dex in [P/Fe], and are indicated in the lower-left corner of the figure. Literature data are taken from \citet{Caffau2011,Jacobson2014,Caffau2016,Maas2017,Afar2018,BocekTopcu2019,Maas2019,Caffau2019,Bocek2020,Sneden2021,Nandakumar2022,Sadakane2022,Catanzaro2025,Ozdemir2025,Matsunaga2026}, and have been homogenized by rescaling the adopted solar abundances of P and Fe to those of \citet{Asplund2009}. We do not include recent studies targeting bulge stars \citep{Barbuy2025,Barbuy2025b,Ernandes2026,Camargo2026}, as these populations are not represented in our sample, nor do we include known P-rich stars from \citet{Masseron2020} and \citet{Brauner2023}, which constitute a distinct class of objects.

Our sample follows the same overall trend established in previous works, with [P/Fe] increasing toward lower metallicities, showing a qualitatively similar metallicity dependence to that commonly observed for $\alpha$-element abundance ratios (e.g., \citealp{Edvardsson1993,Bensby2014,Hayden2015}), although P is not itself an $\alpha$-element. Starting from the metal-rich regime, [P/Fe] rises with decreasing [Fe/H], reaching typical values of $\sim$0.3--0.5 dex around [Fe/H] $\sim -1$. The present survey provides a much denser sampling of this relation over the metallicity range characteristic of the Galactic disk. Figure~\ref{fig:pfe_zoom} provides an enlarged view over $-1.2 \leq [\mathrm{Fe/H}] \leq +0.5$ and $-0.9 \leq [\mathrm{P/Fe}] \leq +0.7$, encompassing the large majority of our sample, and shows the corresponding marginal distributions in [Fe/H] and [P/Fe]. The marginal histograms include detections only, so that their shapes trace the distribution of measured abundances without contributions from upper limits. The [Fe/H] histogram shows that the sample is strongly concentrated around the disk metallicity regime, with the highest density near approximately solar to moderately subsolar metallicities, while the [P/Fe] distribution is centered close to solar values but extends over a broad range toward both enhanced and subsolar abundances. The apparent bimodality in the [P/Fe] distribution is also suggestive of contributions from the chemically distinct thin- and thick-disk populations, which are examined in more detail in Section~\ref{sec:sec4.2}.  Within this regime, the enlarged view reveals a continuous decline in [P/Fe] with increasing metallicity, from enhanced values at the metal-poor end toward approximately solar or mildly subsolar values at solar and super-solar metallicities. The dense sampling also allows the detailed shape and intrinsic dispersion of the [P/Fe]--[Fe/H] relation to be examined more clearly than in previous studies, rather than constraining only its overall normalization.

The much larger number of measurements also makes the dispersion about the mean trend more apparent. At a given metallicity, [P/Fe] spans several tenths of a dex, substantially larger than the median [P/Fe] uncertainty. Measurement uncertainty alone is therefore unlikely to account for the full observed dispersion, suggesting that at least part of the spread may reflect genuine differences among stellar populations. The overall locus defined by the bulk of our sample remains consistent with the trend established by previous studies.

At metallicities around $-1.5 \lesssim [\mathrm{Fe/H}] \lesssim -1.0$ (Figure~\ref{fig:pfe}), although our sample remains sparse, we identify two stars, HIP 79518 ([Fe/H] $= -1.40 \pm 0.02$, [P/Fe] $= -0.07 \pm 0.12$) and HIP 47600 ([Fe/H] $= -1.25 \pm 0.02$, [P/Fe] $= -0.13 \pm 0.20$), which lie near solar [P/Fe]. The larger scatter and small number of stars in this metallicity interval prevent a clear identification of any turnover; however, the low [P/Fe] values seen around $-1.5 \lesssim [\mathrm{Fe/H}] \lesssim -1.0$ suggest that a turnover toward lower abundances, or a flattening around [P/Fe] $\sim 0$, may occur below $[\mathrm{Fe}/\mathrm{H}] \sim -1$, consistent with previous measurements at lower metallicity \citep{Jacobson2014}. Such behavior would differ from the low-metallicity plateau commonly observed in many $\alpha$-element abundance ratios, highlighting that the phenomenological similarity between P and the $\alpha$-elements at higher metallicity does not necessarily extend into the metal-poor regime. Overall, this metallicity regime remains poorly constrained and requires larger samples. In practice, this is challenging, as the P I 10529.52~\AA\ line becomes increasingly weak and eventually undetectable in very metal-poor stars, necessitating ultraviolet spectroscopy to probe P abundances at still lower [Fe/H].

We identify two notable outliers in our sample, HIP 2581 and BD+152 616, which exhibit significantly enhanced [P/Fe] ratios relative to the general trend ([P/Fe] = 0.88 and 1.08, respectively). As such P-rich stars may reflect a distinct enrichment history from the chemically normal population (e.g., \citealp{Brauner2024}), we do not use them when interpreting the main Galactic [P/Fe] trend or the possible low-metallicity flattening discussed above. For HIP 2581, a literature measurement based on low-resolution HST observations reports [Fe/H] $= -0.69$ \citep{Pal2023}, in contrast with our derived value of $-1.34$, while the corresponding $T_{\rm eff}$ and $\log g$ are in good agreement. This discrepancy in metallicity may contribute to its apparent P-rich nature. No independent measurements are currently available for BD+152 616. Both stars represent promising targets for follow-up high-resolution observations to confirm their P enhancement and to investigate their detailed chemical composition in the context of the emerging class of P-rich stars.

In Figure~\ref{fig:pfe}, we also compare our measurements with the scaled Galactic chemical evolution model of \citet{Cescutti2012}, shown as a function of metallicity as a dashed red curve. The model P yields are multiplied by a factor of 2.75, following previous studies \citep[e.g.,][]{Cescutti2012,Maas2022,Nandakumar2022}, to match the observed [P/Fe] abundance level. Overall, the model reproduces the general declining trend of [P/Fe] with increasing metallicity, although differences remain in the detailed shape of the observed relation. 
The large size of our sample therefore constrains not only the normalization of the [P/Fe] trend, but also its metallicity dependence. Uniformly scaling the adopted P yields changes the normalization of the model prediction but not its shape, thus the remaining differences point instead to limitations in the metallicity dependence of the underlying nucleosynthetic yields or to additional P production channels not included in the model. These discrepancies underscore the need for improved nucleosynthetic calculations and chemical evolution models capable of reproducing the detailed structure of the observed [P/Fe]--[Fe/H] relation.

\subsection{[P/Fe] across galactic components} \label{sec:sec4.2}

\begin{figure*}
    \centering
    \includegraphics[width=0.49\textwidth]{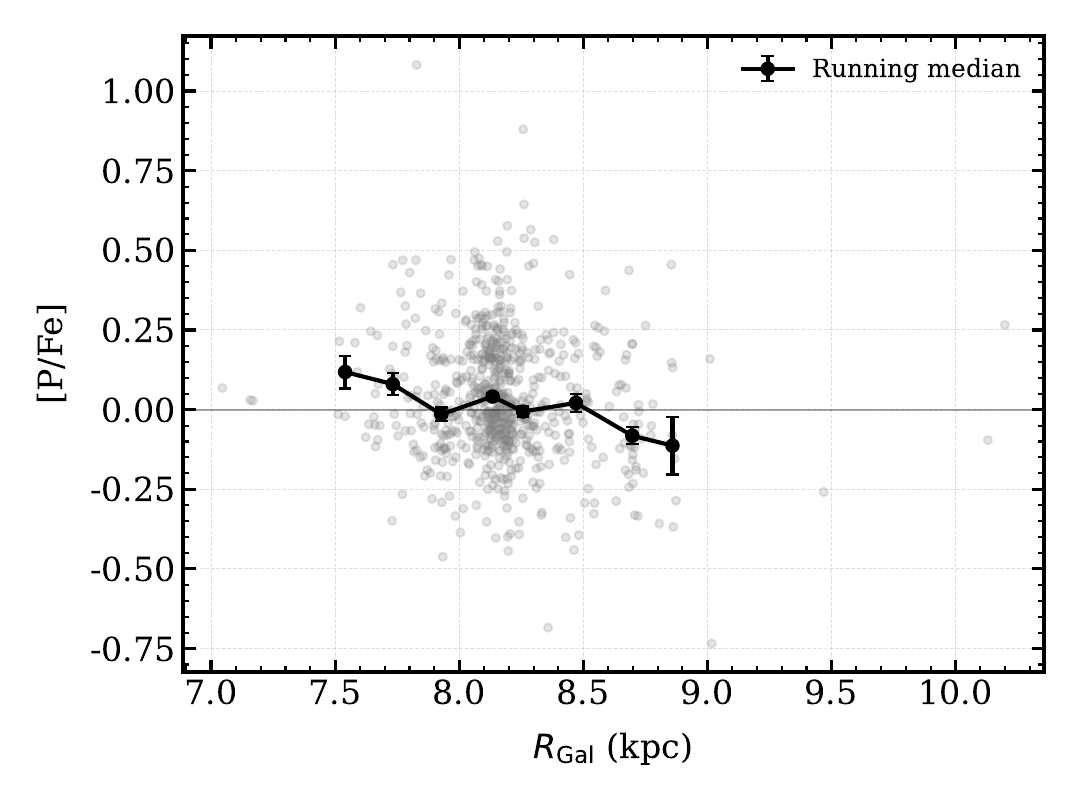}
    \includegraphics[width=0.49\textwidth]{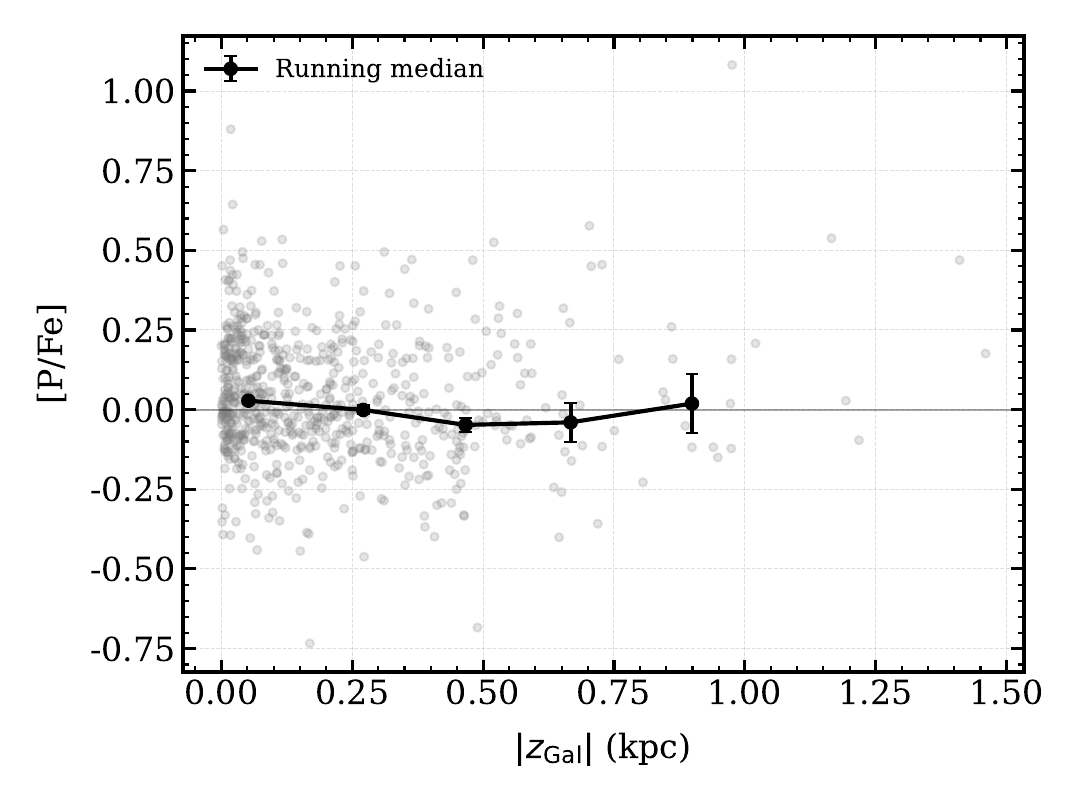}
    \caption{P abundances as a function of Galactocentric position for stars in our sample. 
    \textit{Left:} [P/Fe] as a function of Galactocentric radius ($R_{\rm Gal}$). 
    \textit{Right:} [P/Fe] as a function of vertical height above the Galactic plane ($|z_{\rm Gal}|$). 
    Individual stars are shown as gray points, while black symbols denote the running median in bins of the corresponding coordinate, with error bars representing the standard error of the mean within each bin. }
    \label{fig:pfe_gal}
\end{figure*}

\begin{figure*}
    \centering
\includegraphics[width=0.9\textwidth]{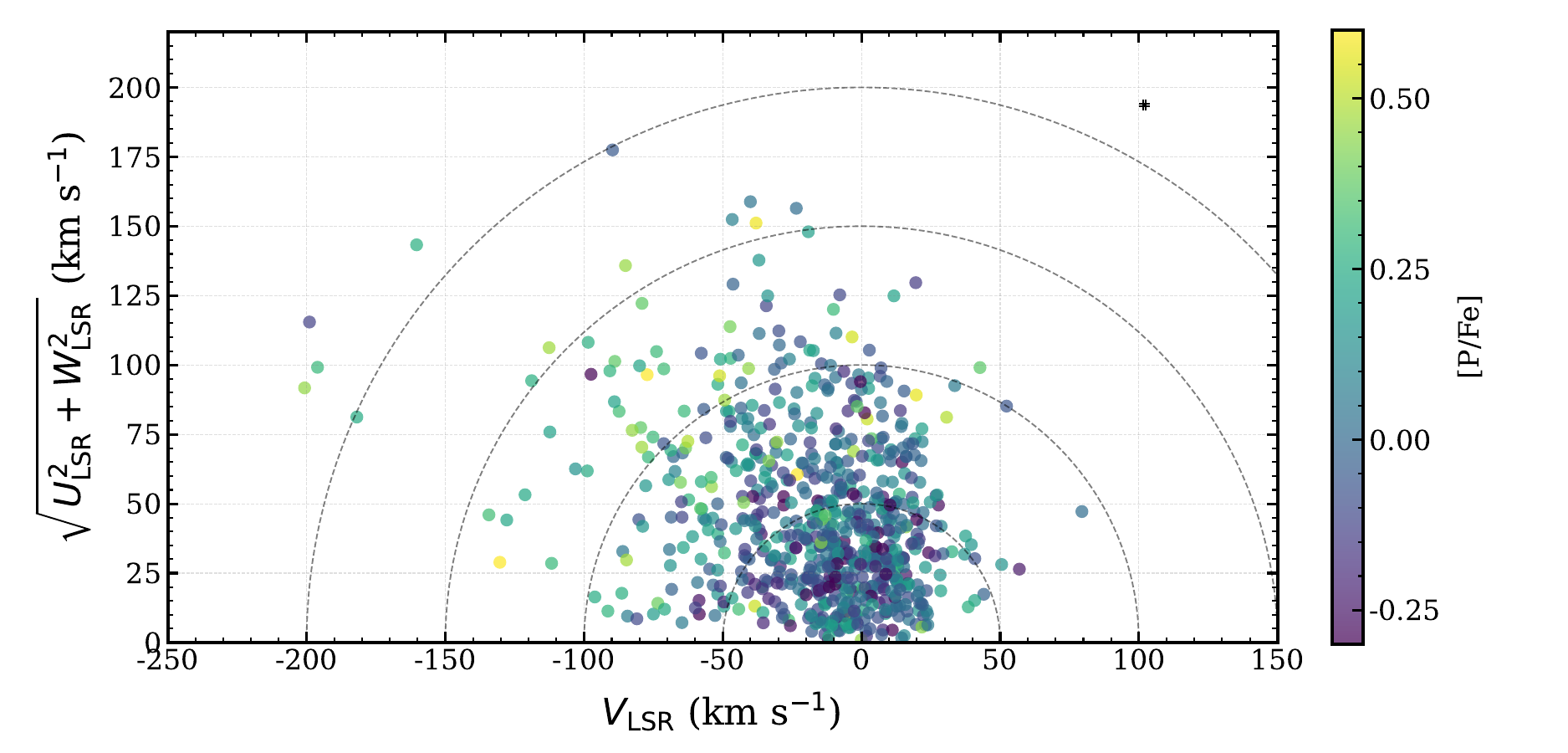}
    \caption{
Toomre diagram showing stellar kinematics, with points color-coded by [P/Fe]. Dashed contours indicate constant total velocity relative to the local standard of rest. The mean uncertainties of the velocity components are shown in the upper-right corner.}
    \label{fig:pfe_toomre}
\end{figure*}

\begin{figure}
    \centering
\includegraphics[width=0.45\textwidth]{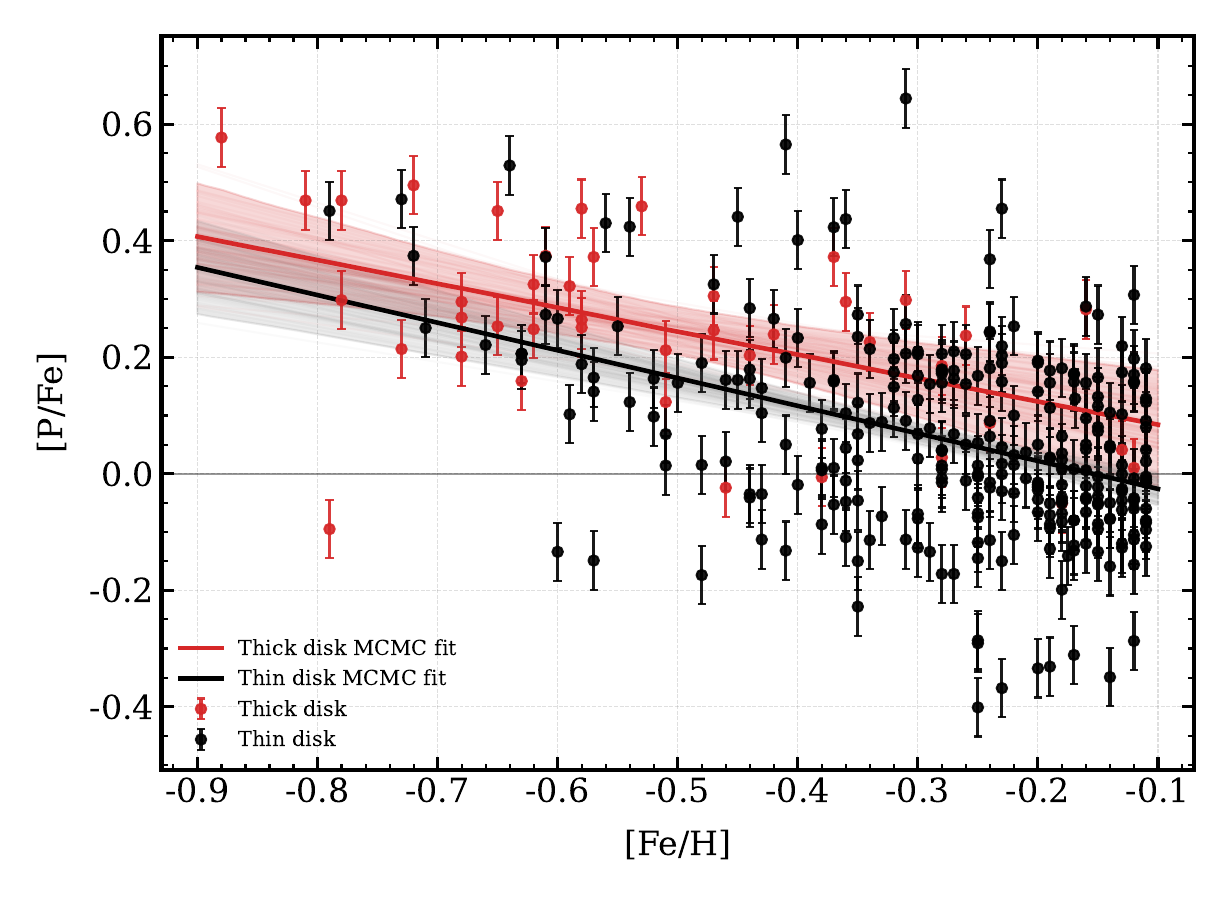}
    \caption{
[P/Fe] as a function of metallicity for kinematically selected thin- (black circles) and thick-disk stars (red circles), restricted to $-0.9 < \mathrm{[Fe/H]} < -0.1$. Points with error bars represent individual measurements and their uncertainties. Solid lines show the median relations obtained from MCMC linear fits for each population, while the shaded regions indicate the corresponding $\sim2\sigma$ (95\%) credible intervals derived from the posterior distributions. Faint lines illustrate 1000 random draws from the posterior to visualize the range of allowed solutions. }
    \label{fig:thick_thin_mcmc}
\end{figure}

The spatial distribution of P abundances is explored in Figure~\ref{fig:pfe_gal}, where we show [P/Fe] as a function of Galactocentric radius ($R_{\rm Gal}$; left panel) and absolute vertical height above the Galactic plane ($|z_{\rm Gal}|$; right panel). Individual stars are shown as gray points, while black symbols denote running medians computed in fixed-width bins of 0.2~kpc, with error bars representing the standard error of the mean within each bin ($\sigma/\sqrt{N}$).

The majority of stars in our sample are located near the solar neighborhood, with a median $R_{\rm Gal}=8.15$~kpc and a full radial extent of $7.04<R_{\rm Gal}<10.20$~kpc, although most stars occupy a substantially narrower range around the solar radius (Figure~\ref{fig:pfe_gal}). A weighted linear fit to the individual [P/Fe] measurements gives a radial gradient of $\Delta[\mathrm{P}/\mathrm{Fe}]/\Delta R_{\rm Gal}=-0.124\pm0.006$~dex~kpc$^{-1}$. This suggests a decrease of [P/Fe] with increasing Galactocentric radius, although the result should be interpreted cautiously given the relatively narrow radial coverage of most of the sample and the substantial scatter about the fitted relation. This radial [P/Fe] gradient has the same sign and is of comparable magnitude to, although somewhat steeper than, the radial metallicity gradients of approximately $-0.06$ to $-0.08$~dex~kpc$^{-1}$ measured for the Galactic disk in previous studies \citep{Eilers2022,Imig2023,Hawkins2023,Hackshaw2024}. By contrast, an APOGEE DR17 open cluster analysis measured a weakly positive radial [Mg/Fe] gradient of $+0.011\pm0.001$~dex~kpc$^{-1}$ \citep{Myers2022}. The radial [P/Fe] gradient measured here is therefore opposite in sign and considerably steeper than the corresponding [Mg/Fe] gradient, although the different stellar tracers and spatial coverage should be considered when making this comparison.


The corresponding vertical fit gives $\Delta[\mathrm{P}/\mathrm{Fe}]/\Delta |z_{\rm Gal}|=+0.052\pm0.007$~dex~kpc$^{-1}$. The nominal positive slope is in the same direction as the vertical behavior generally observed for classical $\alpha$-element ratios, although the considerable scatter and limited number of stars at large $|z_{\rm Gal}|$ warrant caution in interpreting the fitted gradient. Recent APOGEE DR17 and Gaia DR3 results show that [Mg/Fe] increases with maximum vertical orbital height \citep{Horta2024}, while \citet{Zhang2022} measured a vertical [Mg/Fe] gradient of $+0.070\pm0.019$~dex~kpc$^{-1}$ from an open cluster sample. The measured [P/Fe] gradient is of the same sign and broadly comparable in magnitude to this [Mg/Fe] gradient. It differs, however, from the negative vertical metallicity gradient measured by \citet{Hackshaw2024} for the thin disk, $\Delta[\mathrm{Fe}/\mathrm{H}]/\Delta |z_{\rm Gal}|=-0.164\pm0.001$~dex~kpc$^{-1}$. The present sample has a median $|z_{\rm Gal}|=0.125$~kpc and contains relatively few stars at $|z_{\rm Gal}|\gtrsim0.75$~kpc, so the vertical fit is not sufficient to determine whether [P/Fe] changes with height in the broader disk population.

We next examine whether the [P/Fe] behavior is connected to Galactic kinematics. The Toomre diagram in Figure~\ref{fig:pfe_toomre} shows stars color-coded by [P/Fe], with dashed contours indicating constant total velocity relative to the local standard of rest. 
Stars with dynamically hotter kinematics, particularly the 32 stars with $\sqrt{U_{\rm LSR}^2+W_{\rm LSR}^2}\ge75$~km~s$^{-1}$ and $V_{\rm LSR}\le-50$~km~s$^{-1}$, have a median [P/Fe] of $+0.275$ dex and tend to exhibit higher [P/Fe]. Conversely, stars with kinematics typical of the thin disk, characterized by $V_{\rm LSR}$ near zero and low total velocities, $V_{\rm tot}=\sqrt{U_{\rm LSR}^2+V_{\rm LSR}^2+W_{\rm LSR}^2}\lesssim50$~km~s$^{-1}$ \citep[e.g.,][]{Bensby2014}, generally exhibit lower [P/Fe] values. This qualitative behavior suggests that the enhanced-[P/Fe] stars are preferentially associated with older Galactic populations such as the thick disk or halo.

To quantify this separation, we divide the sample into thin- and thick-disk populations using kinematic probability ratios. Following standard criteria \citep[e.g.,][]{Bensby2014}, stars are assigned to the thin disk if $P_{\rm thin}/P_{\rm thick} > 10$ and to the thick disk if $P_{\rm thin}/P_{\rm thick} < 0.1$, while halo-like stars ($P_{\rm halo} \gtrsim 0.1$) and objects with ambiguous disk membership are excluded. This approach closely follows that adopted by \citet{Maas2022}, but is applied here to a significantly larger sample.

We then fit linear relations to the thin- and thick-disk sequences using a Bayesian approach implemented with the \texttt{emcee} package \citep{emcee}. The fits are restricted to the metallicity range $-0.9 < \mathrm{[Fe/H]} < -0.1$, following \citet{Maas2022}. This range covers the metallicities where the thin- and thick-disk sequences are best populated in our sample and avoids the low-metallicity regime where the number of thin-disk stars becomes small. Within this interval, the fitted sample contains 283 thin-disk stars and 42 thick-disk stars. We adopt 50 walkers and run the chains for 10,000 steps, discarding the first 2,000 steps as burn-in. The resulting posterior distributions are used to derive median relations and confidence intervals, shown in Figure~\ref{fig:thick_thin_mcmc}.

The MCMC fits confirm a systematic offset between the two populations, with the thick disk exhibiting higher [P/Fe] at fixed metallicity across the fitted range (Figure~\ref{fig:thick_thin_mcmc}). The inferred linear relations have consistent slopes within uncertainties, with $-0.475^{+0.063}_{-0.062}$ for the thin disk and $-0.406^{+0.108}_{-0.107}$ for the thick disk, while the intercepts differ, at $-0.074^{+0.019}_{-0.019}$ and $0.041^{+0.059}_{-0.059}$, respectively. The median separation between the two sequences is $\sim$0.08~dex, slightly smaller than the $\sim$0.11~dex reported by \citet{Maas2022}. While the separation is statistically significant, the overlap between the populations remains substantial, reflecting both measurement uncertainties and astrophysical scatter. 


Overall, the population-dependent differences identified in our sample suggest that [P/Fe] may vary with Galactic enrichment history, although the magnitude of these differences remains modest and should be interpreted cautiously given the intrinsic scatter and associated uncertainties. The observed trends are consistent with an important contribution from massive star nucleosynthesis, but the present data do not uniquely identify the relative importance of the different channels that may contribute to P production. The relatively large scatter in [P/Fe], together with the need to scale chemical evolution models (e.g., \citealp{Cescutti2012}) to reproduce the observed abundance level, highlights the remaining uncertainties in P nucleosynthesis. In particular, the inability of a simple yield rescaling to reproduce all aspects of the observed [P/Fe]--[Fe/H] relation suggests that revisions to the metallicity dependence of stellar P yields, additional production channels, or both may be required.


\section{Conclusions} \label{sec:sec5}

Phosphorus is an odd-$Z$ element produced in either core-collapse supernovae, massive rotating stars, or C-O shell mergers within stars, and is also of particular interest because of its importance for planetary chemistry and life. However, its Galactic evolution and nucleosynthetic origin remain poorly constrained because reliable stellar P measurements are scarce and have historically been limited to relatively small samples. We have presented the largest homogeneous analysis of P abundances to date, comprising 755 FGK stars observed with the HPF spectrograph and significantly expanding the number of stars with high-resolution near-infrared P measurements. Our analysis uses a spectro-photometric framework applied uniformly across the full sample, combining a newly analyzed set of 601 stars from previous HPF studies with an additional 154 stars from \citet{Maas2022} reprocessed within the same framework. Comparisons with literature compilations and large homogeneous surveys show generally good agreement, supporting the robustness of our parameter estimates and abundance determinations while minimizing methodological biases.

We confirm the previously established increase of [P/Fe] toward lower metallicities, a trend that is qualitatively similar to the metallicity dependence seen in some $\alpha$-element abundance ratios, with typical values of $\sim$0.3--0.5 dex around [Fe/H] $\sim -1$ and substantial intrinsic scatter across the full metallicity range. Around [Fe/H] $\sim -1$, [P/Fe] appears to reach a maximum. We measure a radial [P/Fe] gradient of $-0.124\pm0.006$ dex kpc$^{-1}$ and a vertical gradient of $+0.052\pm0.007$ dex kpc$^{-1}$, although both trends should be interpreted cautiously given the substantial scatter and limited spatial coverage of the sample. Using kinematic classifications, we find that thick-disk stars are systematically enhanced in [P/Fe] relative to thin-disk stars at fixed metallicity by $\sim$0.08 dex, although with significant overlap between the populations. In addition, we identify 2 P-rich candidates that deviate significantly from the main abundance trend and warrant further investigation. Comparisons with Galactic chemical evolution models show that current prescriptions reproduce the general shape of the trend but underpredict the absolute P abundance level, requiring increased yields or additional production channels.

The size and uniformity of this sample enable the [P/Fe]--[Fe/H] relation and its dependence on Galactic population and position to be characterized with substantially improved precision. The tentative low-metallicity turnover, the differences between thin- and thick-disk populations, and the P-rich outliers provide intriguing starting points on the sources and timescales of P enrichment. Comparisons of P with other odd-Z elements, particularly Al and Na, would provide complementary constraints on its nucleosynthetic origin. Future measurements of more metal-poor stars and stars across a wider range of Galactic environments, together with targeted follow-up of the P-rich candidates, will be essential for determining whether the apparent turnover persists and for establishing the nucleosynthetic origin of this rare population.

\section*{Acknowledgements}

We thank the anonymous referee for constructive feedback on the manuscript. DG thanks Carlos Jurado, Zoe Hackshaw and Kaleo Toguchi-Tani for useful discussions that helped improve this work. KH and DG are partially supported by US National Science Foundation (NSF) GLOW grant: 2407975 and NSF AST-2407975. KH acknowledges support from the Wootton Center for Astrophysical Plasma Properties, a U.S. Department of Energy NNSA Stewardship Science Academic Alliance Center of Excellence supported under award numbers DE-NA0003843 and DE-NA0004149, from the United States Department of Energy under grant DE-SC0010623. This work was partly developed at the Scilog - Search For Life sponsored by the Research Corporation and was performed in in part at the Aspen Center for Physics, which is supported by National Science Foundation grant PHY-2210452.

These results are based on observations obtained with the Habitable-zone Planet Finder Spectrograph on the HET. The HPF team acknowledges support from NSF grants AST-1006676, AST-1126413, AST-1310885, AST-1517592, AST-1310875, ATI 2009889, ATI-2009982, AST-2108512, and the NASA Astrobiology Institute (NNA09DA76A) in the pursuit of precision radial velocities in the NIR. The HPF team also acknowledges support from the Heising-Simons Foundation via grant 2017-0494. We acknowledge the Texas Advanced Computing Center (TACC) at The University of Texas at Austin for providing high performance computing, visualization, and storage resources that have contributed to the results reported within this paper.

This research has made use of the NASA Astrophysics Data System Bibliographic Services, and this research has made use of the SIMBAD database, operated at CDS, Strasbourg, France. This publication makes use of data products from the Two Micron All Sky Survey, which is a joint project of the University of Massachusetts and the Infrared Processing and Analysis Center/California Institute of Technology, funded by the National Aeronautics and Space Administration and the National Science Foundation.

Funding for the Sloan Digital Sky Survey IV has been provided by the Alfred P. Sloan Foundation, the U.S. Department of Energy Office of Science, and the Participating Institutions. SDSS acknowledges support and resources from the Center for High-Performance Computing at the University of Utah. The SDSS web site is www.sdss4.org.

SDSS is managed by the Astrophysical Research Consortium for the Participating Institutions of the SDSS Collaboration including the Brazilian Participation Group, the Carnegie Institution for Science, Carnegie Mellon University, Center for Astrophysics | Harvard \& Smithsonian (CfA), the Chilean Participation Group, the French Participation Group, Instituto de Astrofísica de Canarias, The Johns Hopkins University, Kavli Institute for the Physics and Mathematics of the Universe (IPMU) / University of Tokyo, the Korean Participation Group, Lawrence Berkeley National Laboratory, Leibniz Institut für Astrophysik Potsdam (AIP), Max-Planck-Institut für Astronomie (MPIA Heidelberg), Max-Planck-Institut für Astrophysik (MPA Garching), Max-Planck-Institut für Extraterrestrische Physik (MPE), National Astronomical Observatories of China, New Mexico State University, New York University, University of Notre Dame, Observatório Nacional / MCTI, The Ohio State University, Pennsylvania State University, Shanghai Astronomical Observatory, United Kingdom Participation Group, Universidad Nacional Autónoma de México, University of Arizona, University of Colorado Boulder, University of Oxford, University of Portsmouth, University of Utah, University of Virginia, University of Washington, University of Wisconsin, Vanderbilt University, and Yale University.

This work has made use of data from the European Space Agency (ESA) mission
{\it Gaia} (\url{https://www.cosmos.esa.int/gaia}), processed by the {\it Gaia}
Data Processing and Analysis Consortium (DPAC,
\url{https://www.cosmos.esa.int/web/gaia/dpac/consortium}). Funding for the DPAC
has been provided by national institutions, in particular the institutions
participating in the {\it Gaia} Multilateral Agreement. This research has made use of the VizieR catalog access tool, CDS, Strasbourg, France. The original description of the VizieR service was published in \cite{vizier}.

\software{\texttt{Turbospectrum} \citep{turbospectrum}, \texttt{scipy} \citep{scipy}, \texttt{numpy} \citep{numpy}, \texttt{matplotlib} \citep{Matplotlib}, \texttt{pandas} \citep{pandas}, \texttt{astropy} \citep{2013A&A...558A..33A,2018AJ....156..123A,2022ApJ...935..167A}, 
\texttt{emcee} \citep{emcee}, \texttt{muler} \citep{muler}, \texttt{linemake} \citep{linemake}, \texttt{dustmaps} \citep{Green2019}, \texttt{iSpec} \citep{Blanco-Cuaresma2014,Blanco-Cuaresma2019}, \texttt{Goldilocks} HPF Pipeline.}

\bibliography{ref}{}
\bibliographystyle{aasjournalv7}



\end{document}